\documentclass[reprint,superscriptaddress,
nofootinbib,
amsmath,amssymb,
aps,
prd,
floatfix,
notitlepage,
eqsecnum,showpacs
]{revtex4-2}
\usepackage{bm, graphics, epsfig, soul, latexsym}
\usepackage{hyperref}
\usepackage{comment}
\usepackage{subfigure}
\usepackage{graphicx,color}
\usepackage{dcolumn}
\usepackage{mathrsfs}
\usepackage{bigints}
\usepackage{float}
\usepackage{multirow}
\usepackage{footmisc}
\usepackage{placeins}
\usepackage[normalem]{ulem}
\usepackage{booktabs}
\usepackage{tabularx}
\usepackage{xspace}
\usepackage{amsmath}

\begin{document}

\title{Constraining the nature of dark compact objects with spin-induced \\ octupole moment measurement}
\author{Pankaj Saini}
\email{pankajsaini@cmi.ac.in}
\affiliation{Chennai Mathematical Institute (CMI), Siruseri, 603103, India}
\author{N.~V.~Krishnendu}
\email{krishnendu.nv@icts.res.in}
\affiliation{International Centre for Theoretical Sciences (ICTS), Survey No. 151, Shivakote, Hesaraghatta, Uttarahalli, Bengaluru, 560089, India}
\date{\today}

\begin{abstract}
Various theoretical models predict the existence of exotic compact objects that can mimic the properties of black holes (BHs). Gravitational waves (GWs) from the mergers of compact objects have the potential to distinguish between exotic compact objects and BHs. The measurement of spin-induced multipole moments of compact objects in binaries provides a unique way to test the nature of compact objects. The observations of GWs by LIGO and Virgo have already put constraints on the spin-induced quadrupole moment, the leading order spin-induced moment. In this work, we develop a Bayesian framework to measure the spin-induced octupole moment, the next-to-leading order spin-induced moment. The precise measurement of the spin-induced octupole moment will allow us to test its consistency with that of Kerr BHs in GR and constrain the allowed parameter space for non-BH compact objects. For various simulated compact object binaries, we explore the ability of the LIGO and Virgo detector network to constrain the spin-induced octupole moment. We find that LIGO and Virgo at design sensitivity can constrain the symmetric combination of component spin-induced octupole moments of binary for dimensionless spin magnitudes $\sim 0.8$. Further, we study the possibility of simultaneously measuring the spin-induced quadrupole and octupole moments. Finally, we perform this test on selected GW events reported in the third GW catalog. These are the first constraints on spin-induced octupole moment using full Bayesian analysis. 
\end{abstract}
\maketitle
\section{Introduction}
There are theoretical models that predict the existence of exotic compact objects that can mimic the properties of black holes (BHs)~\cite{PhysRevD.94.084031, Giudice:2016zpa,PhysRevD.95.084014,Cardoso:2017cqb,Raposo:2018rjn,Cardoso:2019rvt,Maggio:2021ans}. Black holes and neutron stars (NSs) could be part of a broader category of astrophysical compact objects. The detection of the gravitational wave (GW) signal from the coalescence of compact object binaries by Advanced LIGO~\cite{AdvancedLIGO2010,LIGOScientific:2014pky,KAGRA:2013rdx} and Advanced Virgo~\cite{TheVirgostatus, VIRGO:2014yos} has enabled us to test the nature of these binaries. Till now, the LIGO-Virgo-KAGRA (LVK) Collaboration has reported the detection of $\mathcal{O}(100)$ GW events from the mergers of compact object binaries~\cite{gw150914_detection, GW170817_det,GWTC1-catalog,GW190425detection,GWTC2-catalog,LIGOScientific:2021djp}. The detected GW signals suggest that the binary's constituents are consistent with BHs and NSs in general relativity (GR)~\cite{Krishnendu:2019tjp,GWTC2TGR,GWTC3TGR}.

Boson stars are one example of exotic compact objects that can represent an alternative to BHs~\cite{Mendes:2016vdr, Liebling:2012fv}. Boson stars are gravitational-equilibrium configurations of a massive complex scalar or vector field. Proca stars are a particular class of boson stars that are made up of vector fields~\cite{Sanchis-Gual:2018oui, CalderonBustillo:2020fyi}. Another theoretically proposed exotic objects are gravastars (gravitational vacuum stars) or dark energy stars~\cite{Lobo:2005uf,Chirenti:2007mk,Chirenti:2016hzd,Mottola:2023jxl} which have a de Sitter interior enclosed by a shell of matter. Other alternatives are fermionic stars~\cite{Madsen:1998uh,Alford:2001dt,Weber:2004kj,Gondek-Rosinska:2008zmv,2010PhRvD..81b4012B,DelGrosso:2023trq}, dark matter stars~\cite{Spolyar:2007qv,Tulin:2013teo,Kouvaris:2015rea,Ferrer:2017xwm,Barack:2018yly,Bertone:2019irm}, and multicomponent stars~\cite{Freese:2015mta,Kouvaris:2015rea}.

However, there exist physical phenomena that can generate distinctive characteristics of exotic compact objects, which differ from those of BHs and can be used to distinguish between BH and non-BH compact objects. These physical effects include deformation due to the tidal field of the companion object in a binary system, tidal heating, and deformations due to the spinning motion of binary constituents~\cite{Ryan97b,Poisson:1997ha,Laarakkers:1997hb, Pacilio:2020jza}. For example, the tidal deformability parameter is predicted to be zero for a binary composed of Kerr BHs in an axi-symmetric external gravitational field~\cite{Binnington:2009bb,Damour:2009vw,Chakrabarti:2013lua,Pani:2015hfa,Landry:2015zfa,PhysRevLett.114.151102}. In a recent work, tidal-induced multipole moments are calculated for a Kerr BH with the presence of non-axisymmetric perturbation~\cite{LeTiec:2020spy}. However, exotic compact objects can have non-zero tidal deformability~\cite{Uchikata:2016qku,Cardoso:2017cfl,LeTiec:2020bos}. Parametrizing the gravitational waveform models in terms of these physical effects, many model-agnostic tests have been proposed based on the measurement of spin-induced multipole moments~\cite{Krishnendu:2017shb, Cardoso:2017cfl,Sennett:2017etc}, tidal deformability parameter~\cite{PhysRevD.80.084035,  PhysRevD.80.084018,PhysRevLett.114.151102,Porto:2016zng,JimenezForteza:2018rwr, Abdelsalhin:2018reg,Datta:2019euh,Huang:2020cab,Johnson-Mcdaniel:2018cdu}, and the tidal heating parameter~\cite{PhysRevD.8.1010, PhysRevD.94.084043, Maselli:2017cmm,Datta:2020gem}.

The quasinormal mode spectrum of a perturbed remnant BH after the merger can also be used to probe the nature of compact objects~\cite{Dreyer:2003bv,PhysRevD.73.064030,PhysRevD.85.124056}. For a Kerr BH, all the quasinormal modes are uniquely determined by its mass and spin-angular momentum~\cite{Kamaretsos:2012bs}, while for non-BH compact objects, the quasinormal modes depend on additional parameters~\cite{Berti:2006wq,Gossan:2011ha,Macedo:2013jja,PhysRevD.93.064053}.

In this paper, we focus on the deformations induced due to the spinning motion of compact objects in a binary. To fully understand the nature of compact objects, ideally, one should include all the above-mentioned physical effects along with spin-induced deformations. Studying all the effects simultaneously is beyond the scope of this paper. In particular, we focus on the next-to-leading order spin-induced moment, the {\it spin-induced octupole moment}, which is cubic in the self-spin effect~\cite{Marsat_2014,Bohe:2013cla,Bohe:2012mr,Marsat:2012fn,Marsat:2014xea,Mishra:2016whh}. 

The leading order moment due to the self-spin interaction is the {\it spin-induced quadrupole moment} which is a quadratic self-spin effect and has been studied in great detail~\cite{Krishnendu:2017shb,Krishnendu:2018nqa,Krishnendu:2019tjp,Krishnendu:2019ebd,Saleem:2021vph}. Based on the GW measurements of spin-induced quadrupole moment parameters, Krishnendu et $al.$~\cite{Krishnendu:2017shb} proposed a new method to distinguish between binary black holes (BBHs) and non-binary black holes (non-BBHs). Using the Fisher matrix analysis, they showed that LIGO offers exciting opportunities to constrain the parameter space of non-BBH systems by measuring their spin-induced quadrupole moment. Later in Ref.~\cite{Krishnendu:2018nqa}, this was extended to third-generation (3G) ground-based detectors: Cosmic Explorer (CE)~\cite{LIGOScientific:2016wof,Reitze:2019iox} and Einstein Telescope (ET)~\cite{Punturo:2010zz,Sathyaprakash:2011bh}. Employing the 3G detector sensitivities, Ref.~\cite{Krishnendu:2018nqa} established the possibility of simultaneously measuring spin-induced quadrupole and octupole moments. Similarly, the GW measurements of spin-induced multipole moments using Laser Interferometric Space Antenna (LISA)~\cite{2017arXiv170200786A,Babak:2017tow,2019arXiv190308924A}, and Deci-hertz Interferometer Gravitational wave Observatory (DECIGO)~\cite{Yagi:2013du,Kawamura:2020pcg} established that the space-based detectors could measure the spin-induced multipole moments with an unprecedented accuracy~\cite{Krishnendu:2019ebd}. 

Reference~\cite{Krishnendu:2019tjp} developed a Bayesian framework to carry out the BBH nature test on real GW events based on the spin-induced quadrupole moment. They performed the test on GW151226~\cite{LIGOScientific:2016sjg}, and GW170608~\cite{LIGOScientific:2017vox} and found that these events are consistent with BBH mergers in GR. Subsequently, the LVK Collaboration employed spin-induced quadrupole moment based tests to the second and third GW transient catalogs. The individual event bounds, as well as the combined bounds, provided support for the BBH hypothesis within the allowed statistical confidence~\cite{GWTC2TGR,GWTC3TGR}. More recently, Ref.~\cite{Saleem:2021vph} used the measurement of the spin-induced quadrupole moment to constrain the fraction of non-BBHs in a population consisting of BBHs and non-BBHs.

This study presents the first detailed analysis, utilizing full Bayesian methods, focused on measuring the spin-induced octupole moment of compact binaries. While measuring the spin-induced octupole moment, we fix the spin-induced quadrupole moment to its BH value. Additionally, we explore the possibility of simultaneously measuring the spin-induced quadrupole and octupole moments. Finally, we report the estimates on the spin-induced octupole moment parameter from selected events in the third GW transient catalog (GWTC-3).

\subsection{Spin-induced octupole moment}
The spin-induced quadrupole moment scalar for a compact object can be schematically represented as~\cite{PhysRevD.57.5287} $Q =-\kappa \chi^2 m^3$, where $\kappa$ is the spin-induced quadrupole moment coefficient, $m$ is the mass, and $\chi$ is the magnitude of the dimensionless spin parameter which is defined as $\Vec{\chi}=\Vec{S}/m^2$, where $\Vec{S}$ is the spin angular momentum vector of the compact object. According to the {\it no-hair} conjecture, $\kappa$ takes a unique value for Kerr BHs which is unity $(\kappa_{\rm BH}=1)$~\cite{PhysRevLett.26.331,PhysRevLett.114.151102}. For other compact objects such as NSs, boson stars, and garavastar, the value of $\kappa$ can be different from unity. For NSs, $\kappa$ ranges between $\sim 2$ to $14$ depending on the equation of state~\cite{Laarakkers:1997hb,Pappas:2012qg,Pappas:2012ns}, for spinning boson stars made up of self-interacting massive scalar field, $\kappa$ takes value between $\sim 10$ to $150$~\cite{Ryan97b}. For gravastars, $\kappa$ can take even negative values~\cite{Uchikata:2015yma,PhysRevD.94.064015}. Positive values of $\kappa$ are generally associated with those compact objects that undergo {\it oblate} deformation due to spinning motion, while negative values of $\kappa$ refer to those classes of compact objects which undergo {\it prolate} deformation. Hence a precise measurement of $\kappa$ allows us to constrain the nature of compact objects.

The next-to-leading order moment due to self-spin interactions is the spin-induced octupole moment which can be schematically written as~\cite{Brown:2006pj}
\begin{equation}\label{octupole}
    O = -\lambda \chi^3 m^4 \,,
\end{equation}
where $\lambda$ is the spin-induced octupole moment parameter. Note that the spin-induced quadrupole moment scalar is proportional to quadratic in the component's spin and corrections due to $\kappa$ first appear with the self-spin terms at second post-Newtonian (PN) order in the PN phasing expression~\footnote{In PN phase, a term proportional to $v^{2 n}$ relative to the Newtonian term (proportional to $v^{-5}$) is called the $n$PN order, where $v$ is the characteristic velocity of the system.}. The higher order corrections due to $\kappa$ appear at 3PN and 3.5PN orders. The spin-induced octupole moment has cubic dependence on spin and first appears along with the cubic self-spin terms at 3.5PN order. The value of $\lambda$ is unity for Kerr BHs. For  NSs, $\lambda$ takes value between $\sim 4-30$~\cite{Pappas:2012ns,Laarakkers:1997hb,Pappas:2012qg}, and $\lambda \sim 10-200$ for boson stars consisting of a self-interacting massive scalar field~\cite{Ryan97b}.

The rest of the paper is organized as follows. In Sec.~\ref{sec:waveform model}, we discuss the waveform model and parametrization of the non-BBH nature. Section~\ref{sec:bayesian analysis} explains the basic formalism of the Bayesian analysis. The details of the simulation and results are discussed in Sec.~\ref{sec:results simulation}. The results from GWTC-3 events are shown in Sec.~\ref{sec:real event results}. Section~\ref{sec:conclusions} provides the summary and outlook of the study. Throughout the paper, we use $G=c=1$.

\begin{figure*}
   \centering
   \includegraphics[width=0.99\textwidth]{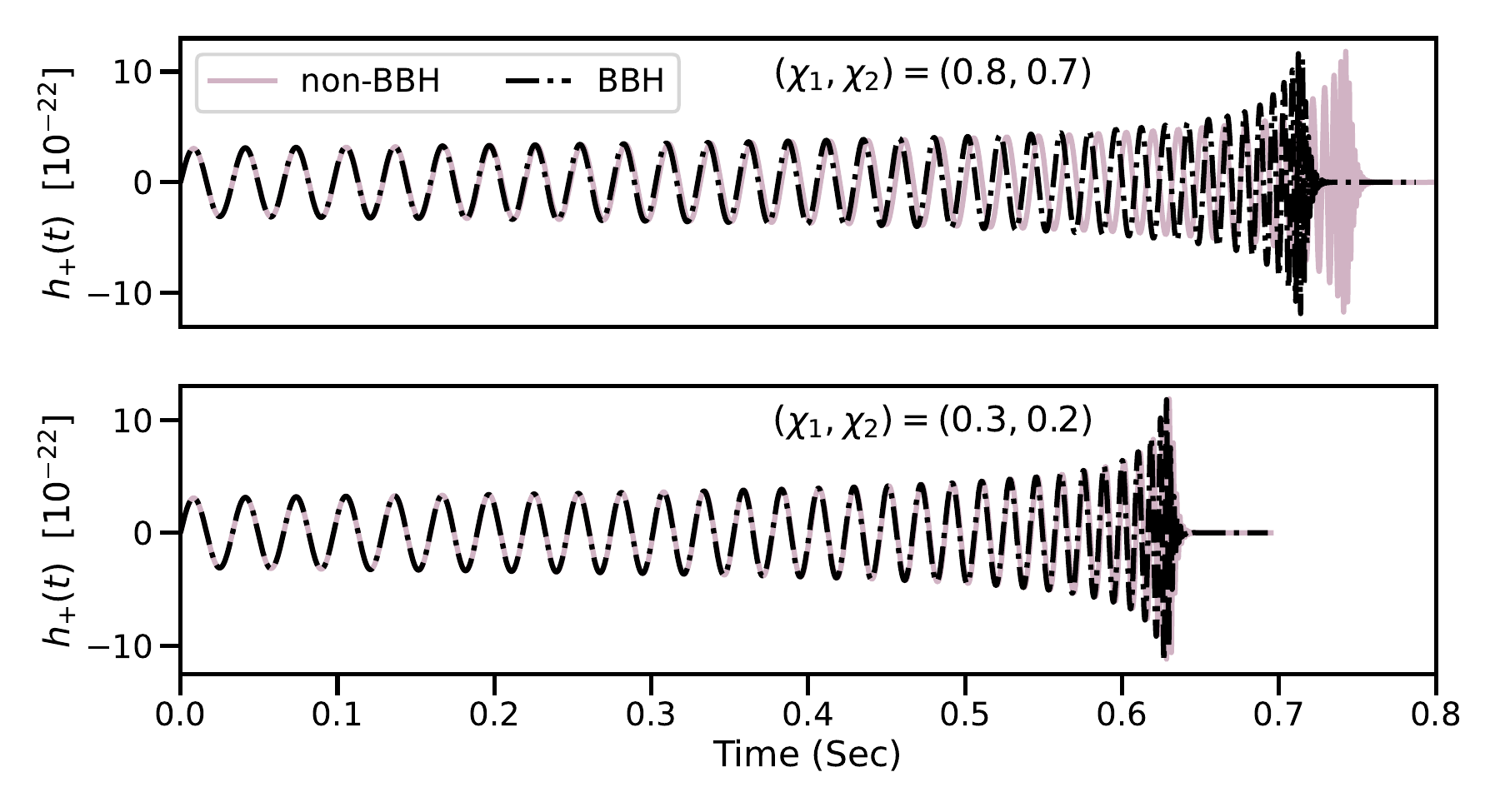}
   \caption{Time-domain gravitational waveform for the BBH (dashed line) and non-BBH (solid line) system with $\text{M}=40 \text{M}_{\odot}$ and $\text{q}=1.2$. The top panel is for high spins $(\chi_1, \chi_2)=(0.8,0.7)$ and bottom panel is for low spins $(\chi_1, \chi_2)=(0.3,0.2)$. Both the spins are aligned with the angular momentum vector (non-precessing). For non-BBH waveforms, the individual spin-induced octupole moment deformations are $\delta\lambda_1 = 100$ and $\delta\lambda_2 = 50$ for primary and secondary components, respectively. We use the {\tt IMRPhenomPv2} waveform model with non-BBH corrections to generate frequency-domain GW signal and take the inverse fast Fourier transform to convert it into the time-domain signal. The BBH and non-BBH waveforms are aligned at time $\text{t}=0$ and the instantaneous frequency at $\text{t}=0$ is $35$ Hz.} \label{fig:waveform}
\end{figure*}

\section{Waveform model and parametrization of non-BBH nature}\label{sec:waveform model}
The inspiral phase of a binary is well described by the PN approximation~\cite{Blanchet:2002av}. The PN expansion is a small velocity expansion in which the inspiral phase is expanded as a series in powers of orbital velocity parameter $v$~\cite{Blanchet:1995ez,Blanchet:1995fg,Kidder:1995zr,Blanchet:2002av,Blanchet:2006gy,Arun:2008kb,Marsat:2012fn,Mishra:2016whh}. The {\it frequency-domain} GW strain, within the PN approximation to GR, can be written as
\begin{equation}\label{waveform}
    \Tilde{h}(f) = \mathcal{A} e^{i\Psi(f)} =\hat{\mathcal{A}} f^{-7/6} e^{i \Psi(f)} \,,
\end{equation}
where $\hat{\mathcal{A}}$ is given by 
\begin{equation}
    \hat{\mathcal{A}} \propto \mathcal{C}(\text{angles}) \frac{\mathcal{M}^{5/6}}{d_L}\,,
\end{equation}
here $\mathcal{M}= (m_1 m_2)^{3/5}/\text{M}^{1/5}$ is the chirp mass of the source, $\text{M} = m_1 + m_2$ is the total mass of the binary, $m_1$, $m_2$ are the component masses of the binary, and $d_L$ is the luminosity distance to the source. Note that $\mathcal{M}$ and $\text{M}$ are the {\it source frame} chirp mass and total mass of the binary. These are converted to the {\it detector frame} (redshifted) chirp mass and total mass by multiplying a factor of $(1+z)$, where $z$ is the redshift to the source. We use the $d_L$ and $z$ relation for a flat universe~\cite{Hogg:1999ad}. The cosmological parameters are taken from the Planck observation~\cite{Planck:2018vyg}. The factor $\mathcal{C}(\text{angles})$ depends on the angles describing the binary sky-position, orientation, and polarization angle of the GW signal.

The phase in Eq.~\eqref{waveform} can be written as
\begin{equation}
\label{SPA}
    \Psi(f) = \phi_c + 2\pi f t_c + \frac{3}{128 \eta v^5}\bigg[ \sum_{i=0}^{7} (\varphi_{i}v^{i} 
      +  \varphi_i^{\rm log}  v^{i} \log v) \bigg] \, ,
\end{equation}
where $\phi_c$ and $t_c$ are the phase and time of coalescence, respectively, $\eta = (m_1 m_2)/\text{M}^2$ is the symmetric mass ratio, and $v= (\pi \text{M} f)^{1/3}$ is the PN orbital velocity parameter. The coefficients $\varphi_i$ and $\varphi_i^{\rm log}$ in Eq.~\eqref{SPA} are the PN coefficients which are the functions of masses and spin angular momenta of the source. The values of these coefficients up to 3.5PN order for quasicircular binaries can be found in Refs.~\cite{Arun:2004hn,Arun:2008kb,Mishra:2016whh,Buonanno:2009zt,Wade:2013hoa,Blanchet:2023bwj,Blanchet:2023sbv}. 

Waveform models for BBHs a priori assume the spin-induced octupole moment parameter to be unity. To parametrize the non-BH nature of spinning compact binaries, we rewrite Eq.~\eqref{octupole} as
\begin{equation}\label{eq:octupole deformation}
    O = -(1+\delta\lambda) \chi^3 m^4\,,
\end{equation}
where $\delta\lambda$ captures the departure from Kerr BH nature. For a Kerr BH, $\delta\lambda=0$. In general, both the compact objects in a binary system can have non-zero $\delta\lambda$. Hence there are two independent spin-induced octupole moment deformation parameters $\delta\lambda_i$, where $i=1,2$ represent the heavier and lighter components of the binary. Because of the strong correlation of $\delta\lambda_i$ with masses and spins, their simultaneous measurement may yield uninformative posteriors (see Refs.~\cite{Krishnendu:2017shb, Krishnendu:2018nqa} for the spin-induced quadrupole moment measurement). Hence following~\cite{Krishnendu:2017shb}, we define the symmetric and antisymmetric combinations of the individual spin-induced octupole moment deformation parameters as 
\begin{equation}
  \delta\lambda_s = \frac{(\delta\lambda_1 + \delta\lambda_2)}{2}\,, \;\; \delta\lambda_a = \frac{(\delta\lambda_1 - \delta\lambda_2)}{2} \,.   
\end{equation} 
We measure only $\delta\lambda_s$ and assume $\delta\lambda_a=0$, which amounts to assuming $\delta\lambda_1 = \delta\lambda_2 = \delta\lambda_s$, which means the two compact objects are of the same nature and hold well for BBHs. If a binary is composed of a non-BH compact object, this would violate this condition leading to a non-zero value of $\delta\lambda_s$, revealing the presence of a non-BH compact object in the binary. 

Similarly, the GW phase can be simultaneously parametrized in terms of the symmetric and antisymmetric combination of deformations due to spin-induced quadrupole moment ($\delta\kappa_s, \delta\kappa_a$). Ideally, one should measure both $\delta\kappa_s$ and $\delta\lambda_s$ simultaneously. However, the simultaneous measurement would yield weaker constraints. As a first step, we assume the BBH value for spin-induced quadrupole moments ($\delta\kappa_s=0, \delta\kappa_a=0$) and focus on the measurement of $\delta\lambda_s$. As a null test, this would independently check the consistency of the spin-induced octupole moment with the BBH nature in GR. Next, we vary both $\delta\kappa_s$ and $\delta\lambda_s$ and obtain the simultaneous constraints on $\delta\kappa_s$ and $\delta\lambda_s$. The expressions for PN phasing with explicit $\kappa_s$ and $\lambda_s$ dependence are given in Eq.(0.5) of Ref.~\cite{Krishnendu:2017shb}. 

We use the {\tt IMRPhenomPv2}~\cite{Ajith:2009bn,Hannam:2013oca,Husa:2015iqa} waveform approximant which is implemented in {\tt LALSimulation} \cite{lalsuite}. {\tt IMRPhenomPv2} is a frequency-domain phenomenological waveform model that describes the inspiral-merger-ringdown of a BBH coalescence in GR. The inspiral part agrees with the PN phasing and merger-ringdown part is calibrated using numerical relativity waveforms. {\tt IMRPhenomPv2} is a single-spin precessing waveform and accounts only for the dominant modes $(l,m)=(2,\pm 2)$. The deformations due to spin-induced quadrupole moment ($\delta\kappa_s, \delta\kappa_a$) have already been implemented up to 3PN order~\cite{Krishnendu:2019tjp}. We add the leading order octupole deformations ($\delta\lambda_s, \delta\lambda_a$) at 3.5PN order in the inspiral part of the waveform. Additionally, we include the $\delta\kappa_s$ and $\delta\kappa_a$ corrections at 3.5PN order. We make necessary modifications to {\tt bilby}~\cite{Ashton:2018jfp} and {\tt bilby\textunderscore pipe}~\cite{Romero-Shaw:2020owr} for parameter estimation. We perform Bayesian analysis using {\tt bilby} to obtain posterior on $\delta\lambda_s$, $\delta\kappa_s$, and other binary parameters. For sampling over the intrinsic and extrinsic parameters, we use {\tt dynesty}~\cite{speagle2020dynesty} sampler that uses the nested sampling algorithm~\cite{2004AIPC..735..395S}.

In Fig.~\ref{fig:waveform}, we plot the time-domain BBH ($\delta\lambda =0 $) and non-BBH ($\delta\lambda \neq 0$) waveform for high spins $(\chi_1, \chi_2) = (0.8, 0.7)$ and low spins  $(\chi_1, \chi_2) = (0.3, 0.2)$. Spins are aligned with the orbital angular momentum of the binary (non-precessing). The total mass and mass ratio are chosen to be $\text{M}=40 \text{M}_\odot$ and $\text{q}=1.2$, respectively. The primary and secondary components in a non-BBH system have $(\delta\lambda_1, \delta\lambda_2) = (100, 50)$. The time-domain GW signal is generated by taking inverse fast Fourier transform of the frequency-domain waveform {\tt IMRPhenomPv2}. The waveform for $(\chi_1, \chi_2)= (0.8,0.7)$ is longer than the waveform with $(\chi_1, \chi_2)= (0.3,0.2)$. Notice that the dephasing between BBH and non-BBH waveforms increases as the magnitude of spins increases. In other words, the waveform has more signature of non-BBH nature for high values of spins. 

Note that the test relies on the assumption that our parametric waveform model captures deviation from the Kerr nature of BHs. A non-BBH can have different properties such as additional fields, modes, and frequency evolution. Therefore, any deviation from the Kerr nature needs further investigation employing a non-BBH waveform model specific to a particular class of non-BH compact object to identify the true nature of the compact binary.

\begin{table*}
\centering
\renewcommand{\arraystretch}{1.32}
\setlength{\tabcolsep}{8pt}
\caption{Summary of the properties of simulated binary black holes. We choose three detector frame total masses keeping the mass ratio $\text{q}=m_1/m_2$ and SNR to be fixed at $1.2$ and $40$, and three mass ratios for fixed total mass $\text{M}=40\text{M}_{\odot}$ and $\text{SNR}=40$. For $\text{M}=40\text{M}_{\odot}$ and $\text{q}=1.2$, we choose three network SNRs by varying the luminosity distances $d_L$ as shown in the bottom row. The sky-location and orientation are chosen randomly but are the same for all systems.  For each system, we choose three spin combinations $(\chi_1, \chi_2):$ $(0.3, 0.2)$, $(0.5, 0.4)$, and $(0.8,0.7)$.} 
\vspace{2mm}
\begin{tabular}{c|c|c|c|c|c|c|c|c|c}
\hline
\hline
& \multicolumn{3}{c|}{$\text{q}=1.2$} &  \multicolumn{3}{c|}{$\text{M}=40 M_{\odot}$} &  \multicolumn{3}{c}{$\text{M}=40 M_{\odot}$} \\
&  \multicolumn{3}{c|}{$\text{SNR}=40$} &  \multicolumn{3}{c|}{$\text{SNR}=40$} &  \multicolumn{3}{c}{$\text{q}=1.2$} \\ 
\hline 
& \multicolumn{3}{c|}{\textbf{M}} &  \multicolumn{3}{c|}{\textbf{q}} &  \multicolumn{3}{c}{\textbf{SNR}} \\ 
\hline
& $20 M_{\odot}$ & $40 M_{\odot}$ & $70 M_{\odot}$ & $1.2$ & $3$ & $5$ & $40$ & $80$ & $120$ \\
\hline 
$d_L$ [Mpc] & $450$ & $840$ & $1225$ & $840$ & $700$ & $570$ & $840$ & $420$ & $280$\\ 
\hline
\hline
\end{tabular}
\label{tab:simulation}
\end{table*}

\section{A review of Bayesian parameter estimation}\label{sec:bayesian analysis}
We use the Bayesian parameter inference~\cite{Veitch:2009hd,Veitch:2014wba,thrane2019introduction} to obtain the posterior probability distribution on model parameters $\bm{\theta}$ given the data $d$ from GW observations. According to the Bayes theorem, the posterior distribution on $\bm{\theta}$, given the GW data $d$, is given by
\begin{equation}\label{posterior}
    p(\bm{\theta}|d) = \frac{\mathcal{L}(d|\bm{\theta}) \mathcal{\pi}(\bm{\theta})}{p(d)} \,,
\end{equation}
where, $\mathcal{L}(d|\bm{\theta})$ is the likelihood function of the data $d$ given the model parameters $\bm{\theta}$, and $\mathcal{\pi}(\bm{\theta})$ is the prior probability for $\bm{\theta}$. The term in the denominator $p(d)$ is a normalization factor which is obtained by marginalizing the likelihood over the prior and is called the “evidence”,
\begin{equation}\label{evidence}
  p(d) = \int \mathcal{L}(d|\bm{\theta}) \mathcal{\pi}(\bm{\theta}) d\bm{\theta}\,.
\end{equation}
Under the Gaussian noise assumption, the likelihood function is given by
\begin{equation}
    \mathcal{L}(d|\bm{\theta}) \propto  \exp \bigg(-\frac{1}{2} (d-h(\bm{\theta})\big|d-h(\bm{\theta}))\bigg) \,.
\end{equation}
where $h(\bm{\theta})$ is a template for the gravitational waveform given by $\bm{\theta}$. The inner product $(\cdots|\cdots)$ is defined as 
\begin{equation}\label{innerproduct}
    (A|B) = 2 \int_{f_{\rm low}}^{f_{\rm high}}\frac{\Tilde{A}^{*}(f) \Tilde{B}(f) + \Tilde{A}(f)\Tilde{B}^{*}(f) }{S_{n}(f)}\, df,
\end{equation}
where $S_{n}(f)$ is the one-sided noise power spectral density (PSD) of the detector and $\ast$ represents the complex conjugation. The limits of integration in Eq.~\eqref{innerproduct} are fixed by the detector sensitivity and the source properties.

The model parameters $\bm{\theta}$ consist of BH and non-BH (NBH) parameters $h(\bm{\theta}) \equiv h(\Vec{\theta}_{\rm BH},\Vec{\theta}_{\rm NBH})$. The dynamics of a BBH in a quasicircular orbit is described by $15$ parameters namely, component masses, component spin vectors, sky-position and orientation of the binary, and phase at the detector. The $\delta\kappa_s$ and $\delta\lambda_s$ are the non-BBH parameters that are introduced in the waveform model to capture the non-BBH nature of compact object binary. 

The one-dimensional posterior on $\Vec{\theta}_{\rm NBH}$ can be obtained by marginalizing over $\Vec{\theta}_{\rm BH}$
\begin{equation}\label{nonBH}
   p(\Vec{\theta}_{\rm NBH}|d) = \int p(\bm{\theta}|d) d\Vec{\theta}_{\rm BH}\, .
\end{equation}
Using these Bayesian analysis methods and the waveform model discussed in Sec.\ref{sec:waveform model}, we obtain posteriors on spin-induced multipole moments.

\section{Constraints on non-BBH nature from simulated events}
\label{sec:results simulation}
\subsection{Simulation setup}
\label{simulation}
We study the possible measurement uncertainties in the spin-induced octupole moment deformation parameter ($\delta\lambda_s$) for various binary systems with different intrinsic and extrinsic properties. 

\textbf{Network configuration:} We consider a network of three detectors (HLV): two advanced LIGO detectors at Hanford (H) and Livingston (L) and an advanced Virgo detector (V) in Italy. For noise PSD, we use the design sensitivity curve of LIGO and Virgo detectors~\cite{H1L1V1Dcc}. All injections are zero-noise injections to avoid any bias in the estimated parameters due to particular noise realizations. It is equivalent to averaging over many noise realizations. The detector noise only contributes to the likelihood. We use a lower cut-off frequency of $20$ Hz for all three detectors and the upper cutoff frequency is the Nyquist frequency which is $0.875\times(f_s/2)$, where $f_s$ is the sampling frequency~\cite{LIGOScientific:2021djp}.

\textbf{Masses:} We consider three detector frame total masses:  $20\text{M}_{\odot}$, $40\text{M}_{\odot}$, and $70\text{M}_{\odot}$. The mass ratio for all three systems is fixed to be $1.2$. The Network SNR is fixed to be $40$ by varying $d_L$ (see Table~\ref{tab:simulation}). 

\textbf{Mass ratios:} Three different mass ratios $(q=m_{1}/m_{2})$ are chosen to be $1.2$, $3$, and $5$. The total mass is fixed to be $\text{M}=40\text{M}_{\odot}$. The network SNR is fixed to $40$. 

\textbf{Spins:} For each binary system mentioned in Table~\ref{tab:simulation}, we choose three combinations of dimensionless spin magnitudes ($\chi_1$, $\chi_2$): $(0.3,0.2)$, $(0.5,0.4)$ and $(0.8,0.7)$ which represent low, moderate, and high spins, respectively. 
The precession angles~\footnote{The angles between spins and the orbital angular momentum vector $(\theta_1, \theta_2)$, the difference between total and orbital angular momentum azimuthal angles  $(\phi_{\rm JL})$, and the difference between the azimuthal angles of the individual spin vector projections onto the orbital plane  $(\phi_{12})$.} are chosen randomly but the same for all systems.  The injected values of these angles are $(\theta_1, \theta_2) = (0.5, 1)$, $\phi_{\rm JL} = 0.3$, and $\phi_{12} = 1.7$.

\textbf{$\delta\lambda_s$ parameter:} To investigate the detectability of the non-BBH nature in the detected signal, we simulate binaries with different values of $\delta\lambda_s=\{-100, -50, -15, 15, 50, 100\}$. 

\textbf{Network SNRs:} For $\text{M}=40\text{M}_{\odot}$ and $\text{q}=1.2$, we choose three network SNRs, $40$, $80$, and $120$. The corresponding $d_L$ is mentioned at the bottom row of Table~\ref{tab:simulation}. Note that there is a slight variation in the SNR due to a change in the magnitude of spins of the binary. For higher values of spins, the SNR is larger. However, the difference in SNR due to the choice of different spins is $\lesssim 5\%$.

\textbf{Extrinsic parameters:}  The sky-location $(\alpha, \delta)$ and orientation  $(\theta_{\rm JN})$ are chosen randomly but are the same for all systems. The values of these parameters are $\alpha=1.375$, $\delta=1.21$, $\theta_{\rm JN}= 0.4$. The changes in sky-location and orientation can affect the $\delta\lambda_s$ estimates only through the SNR.

\textbf{Priors:} Binary black hole injections are generated with $\delta\kappa_s=0$ and $\delta\lambda_s=0$. We use flat priors on $\delta\lambda_s$ in the range $[-500,500]$, which is the same for $\delta\kappa_s$. The priors on component masses are chosen to be uniform in the range $[5,100] \text{M}_{\odot}$ \footnote{For $\text{M}=40\text{M}_\odot$ with $\text{q}=3$ and $\text{q}=5$, the lower limit on component mass prior is chosen to be $1 \text{M}_{\odot}$ instead of $5 \text{M}_{\odot}$ to avoid any bias on the secondary component due to restricted prior range.}. The priors on $(\chi_1, \chi_2)$ are uniform in the range $[0,0.99]$. 

\subsection{Constraints on $\delta\lambda_{s}$}\label{simulation results}
\begin{figure}
\centering
   \subfigure{\includegraphics[width=0.49\textwidth]{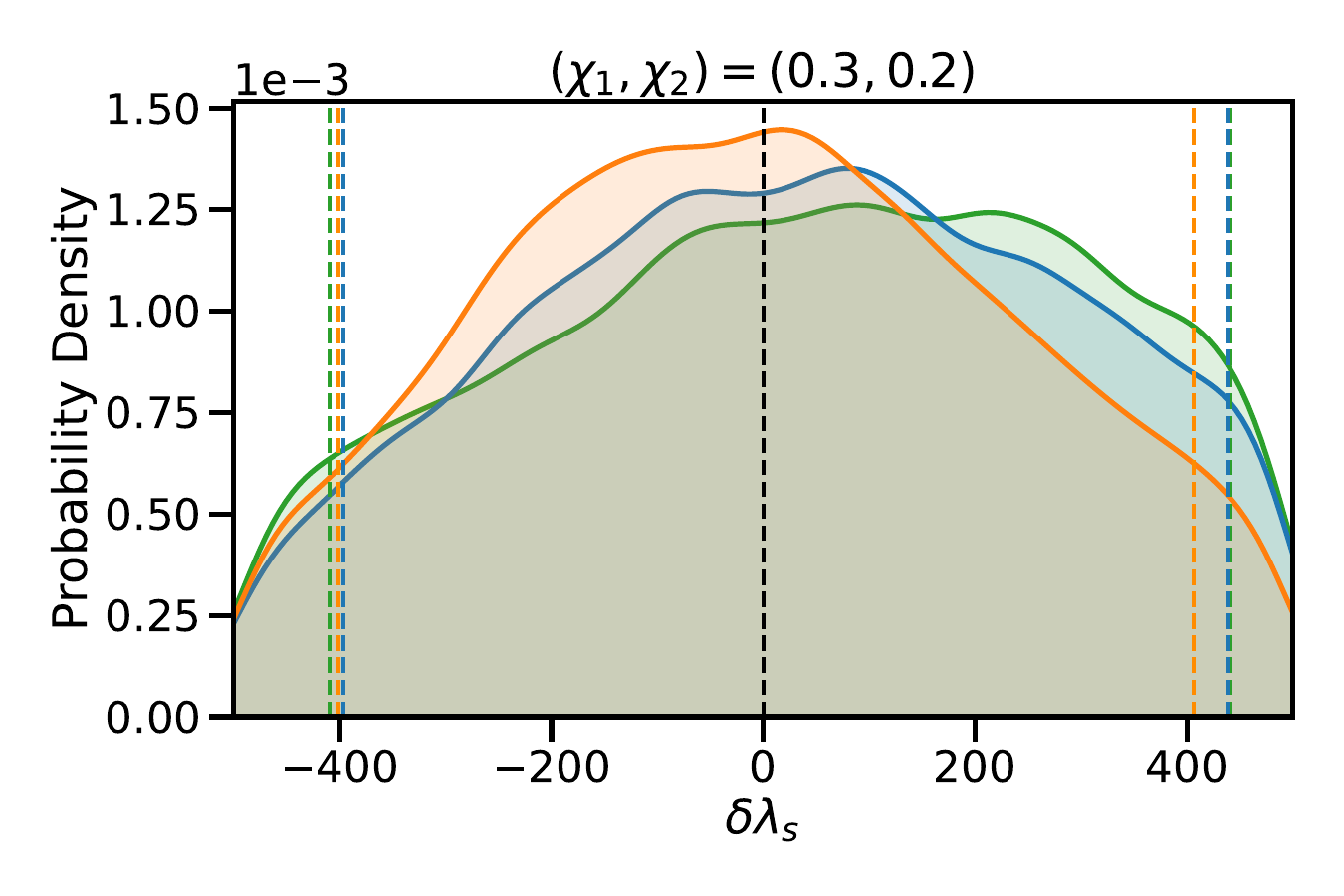}} \vspace{-0.7cm}
   
   \subfigure{\includegraphics[width=0.49\textwidth]{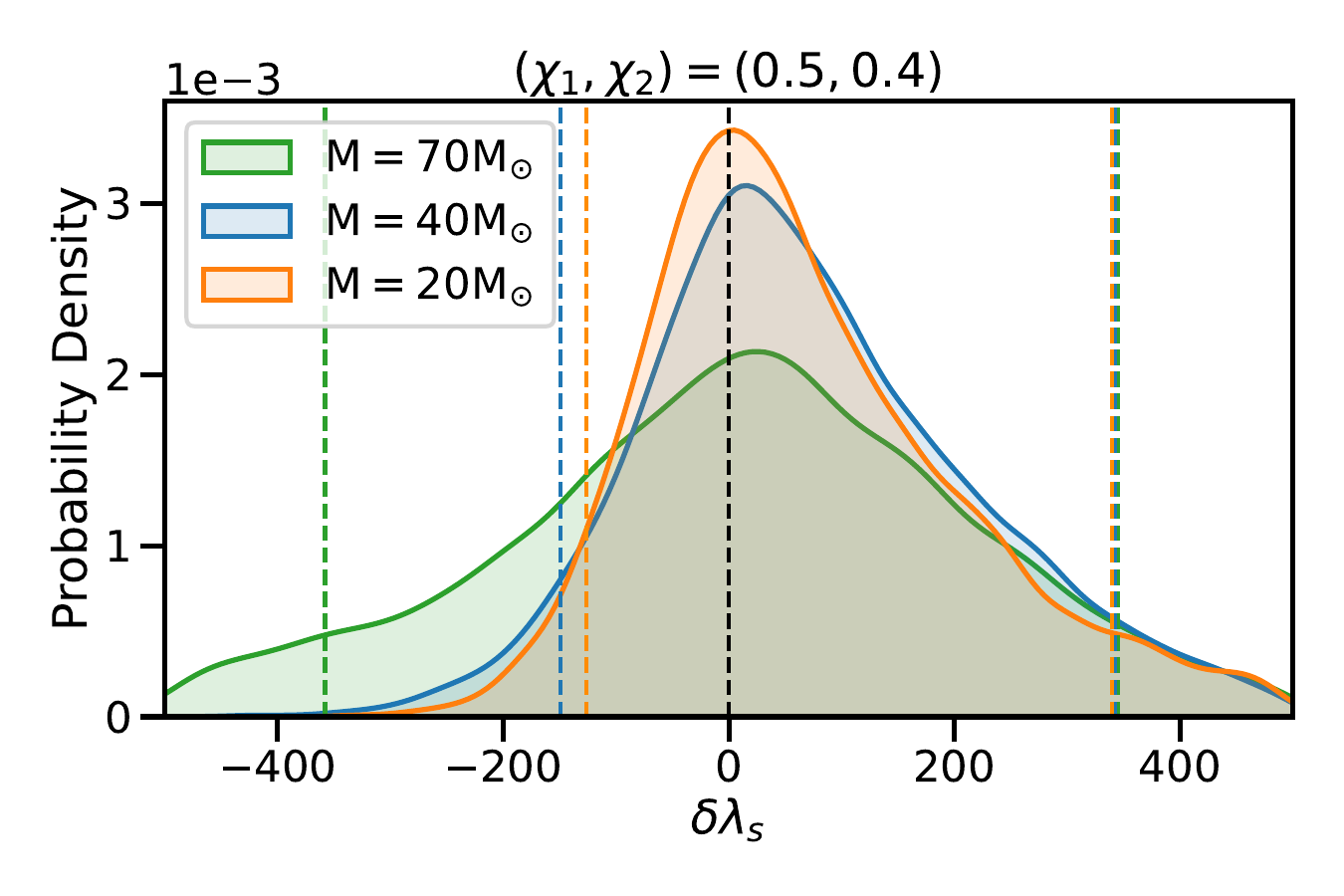}}\vspace{-0.7cm} 
   
   \subfigure{\includegraphics[width=0.50\textwidth]{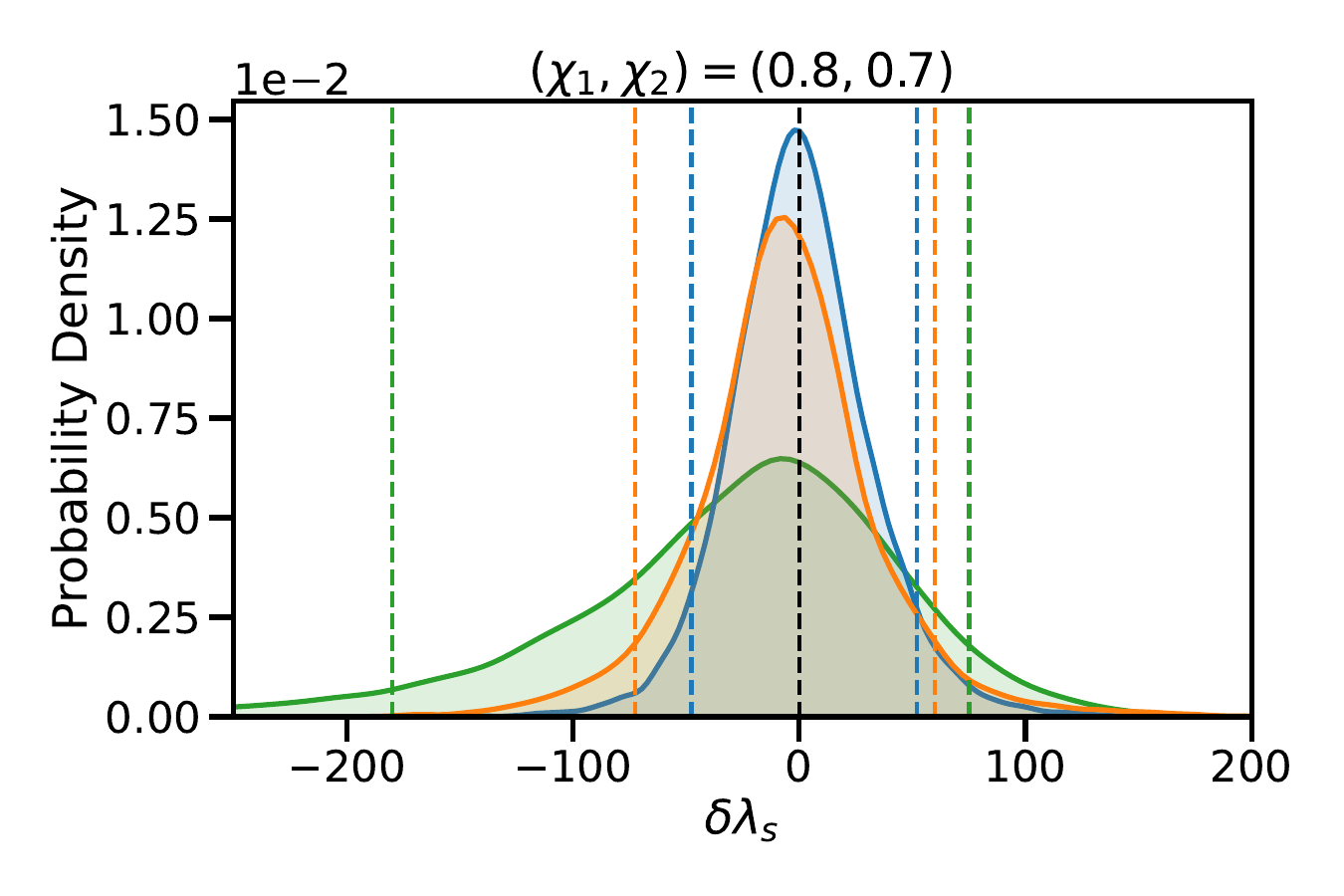}}
\caption{Posterior probability distribution on spin-induced octupole moment parameter $\delta\lambda_s$ for three different values of spins: low spin ($0.3,0.2$), moderate spin ($0.5, 0.4$), and high spin ($0.8, 0.7$). Three different colors in each plot represent three different masses. The $90\%$ credible interval is marked by the vertical lines in the respective colors. The black vertical line represents the BBH value ($\delta\lambda_s=0$). The mass ratio is fixed to be $\text{q}=1.2$. The SNR is kept the same for all the masses by varying the distances (see Table~\ref{tab:simulation}). Sky-location, orientation, and precession angles are chosen randomly but the same for all.} 
\label{fig:mass}
\end{figure}

\begin{figure}
\centering
   \subfigure{\includegraphics[width=0.49\textwidth]{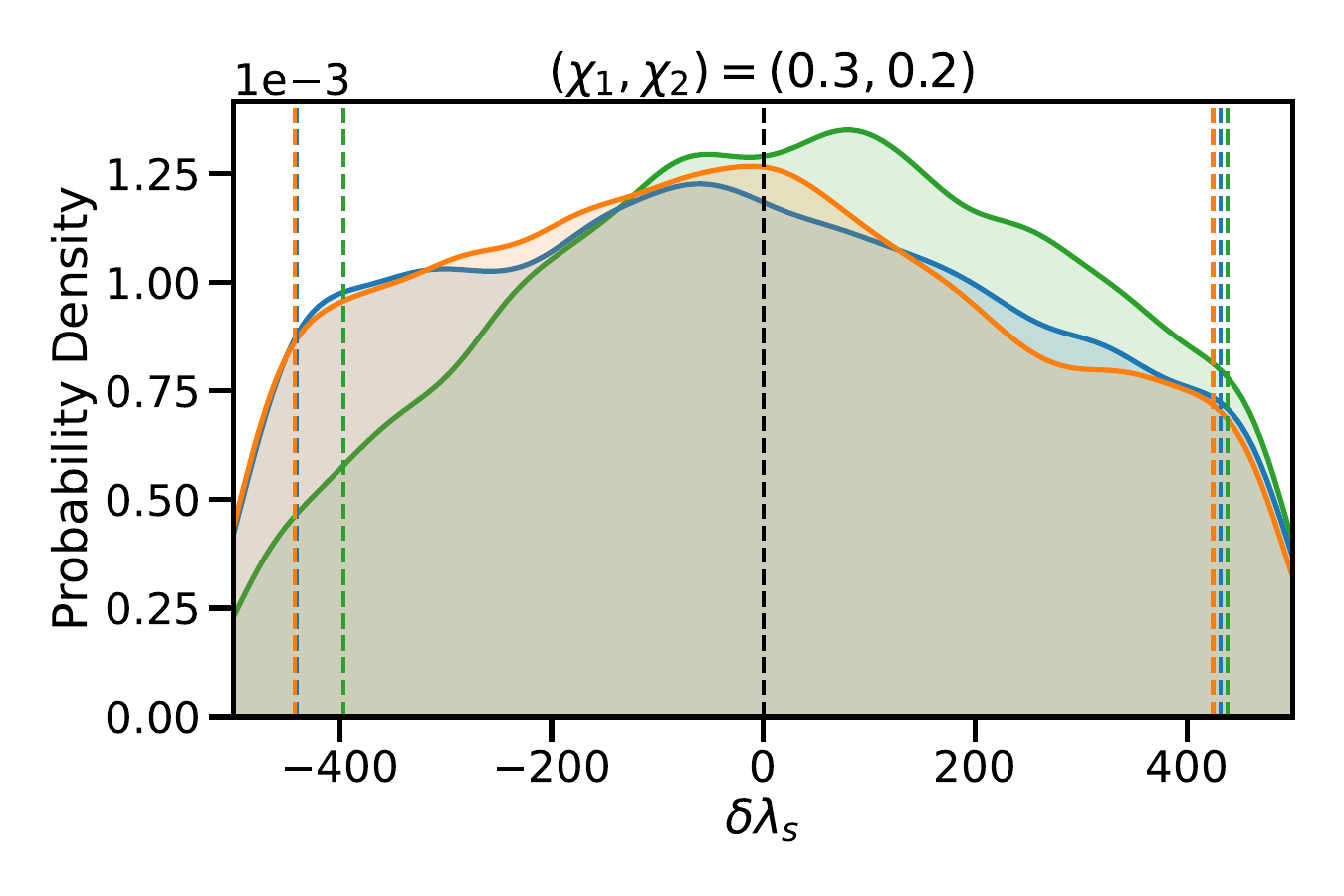}}\vspace{-0.7cm}
   
  \subfigure{\includegraphics[width=0.49\textwidth]{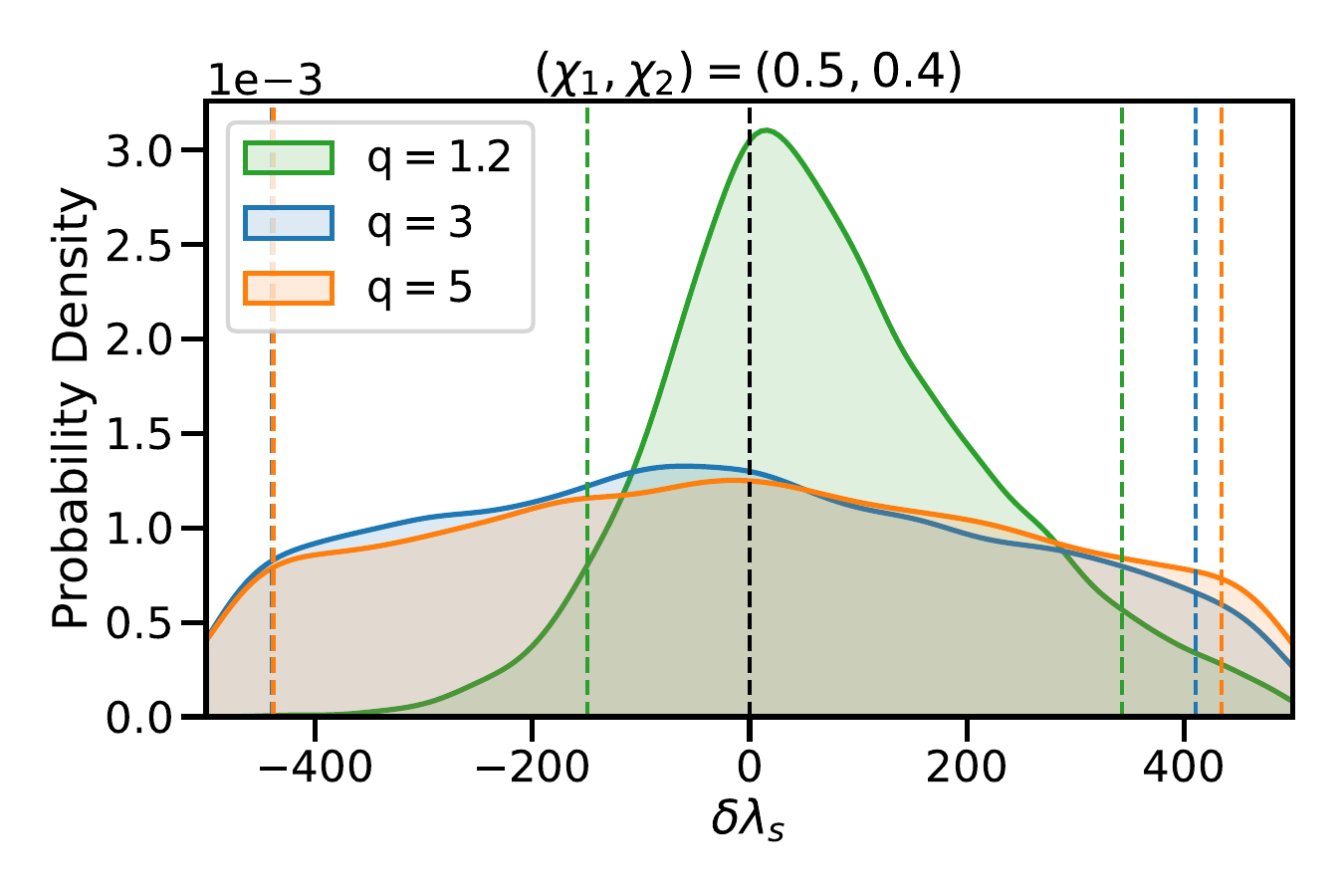}}\vspace{-0.7cm}
  
\subfigure{\includegraphics[width=0.49\textwidth]{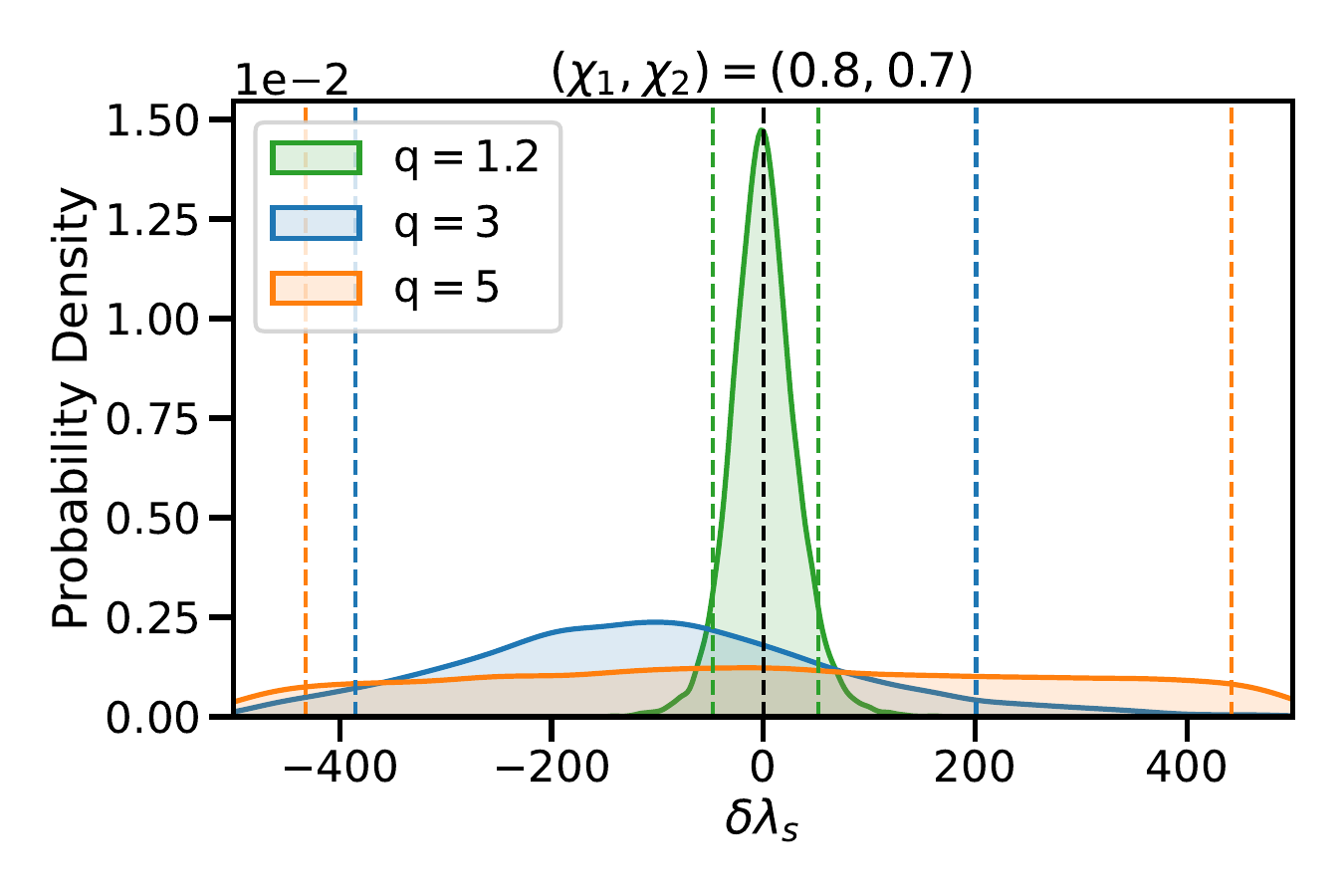}}  
\caption{Posterior probability distribution of $\delta\lambda_s$ for three different mass ratios ($\text{q}=1.2, 3, 5$). The total mass is fixed to be $\text{M}=40\text{M}_{\odot}$. Other parameters are the same as Fig.~\ref{fig:mass}.}
\label{fig:massratio}
\end{figure}

\begin{figure}
\centering
\subfigure{\includegraphics[width=0.49\textwidth]{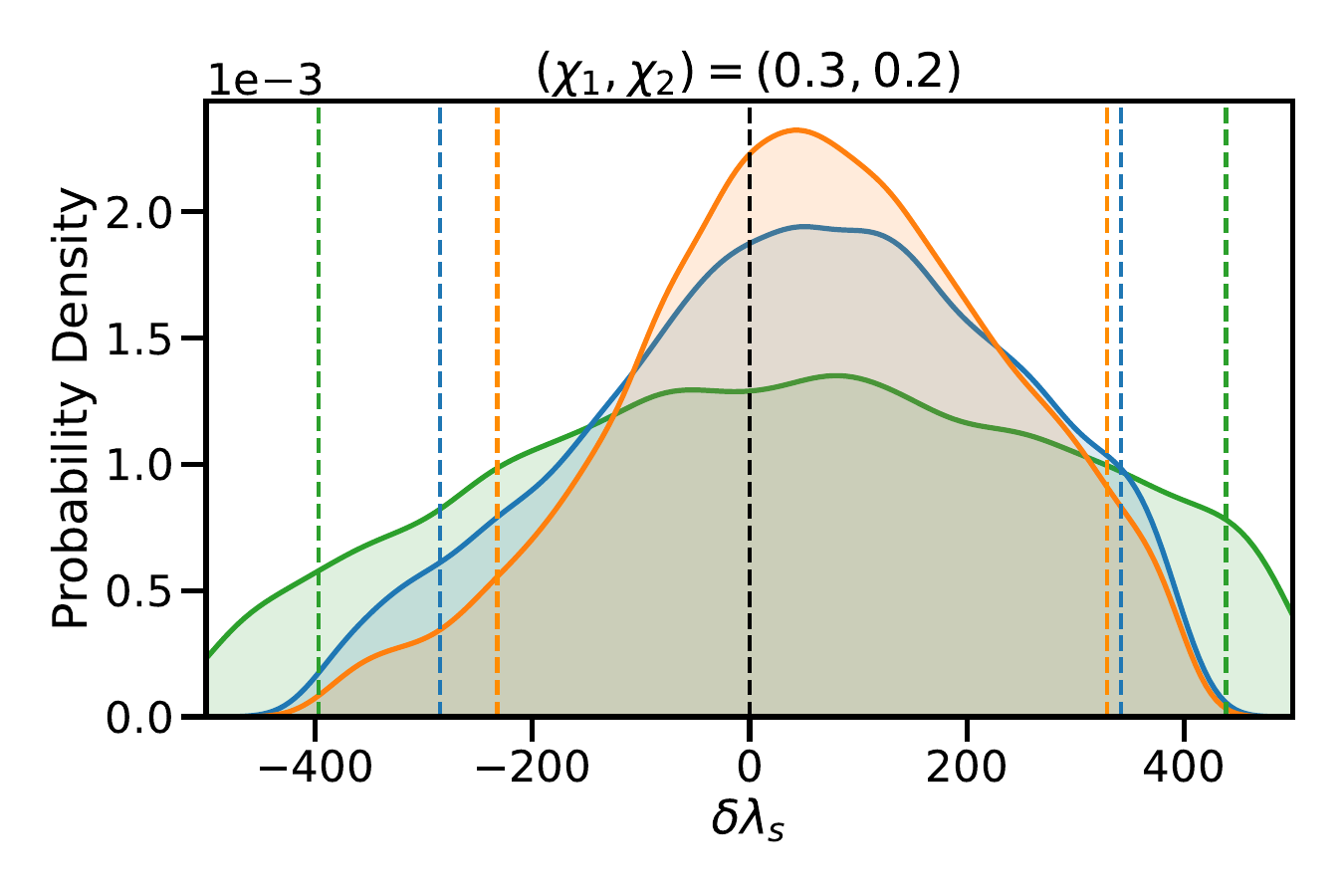}}\vspace{-0.7cm}

\subfigure{\includegraphics[width=0.49\textwidth]{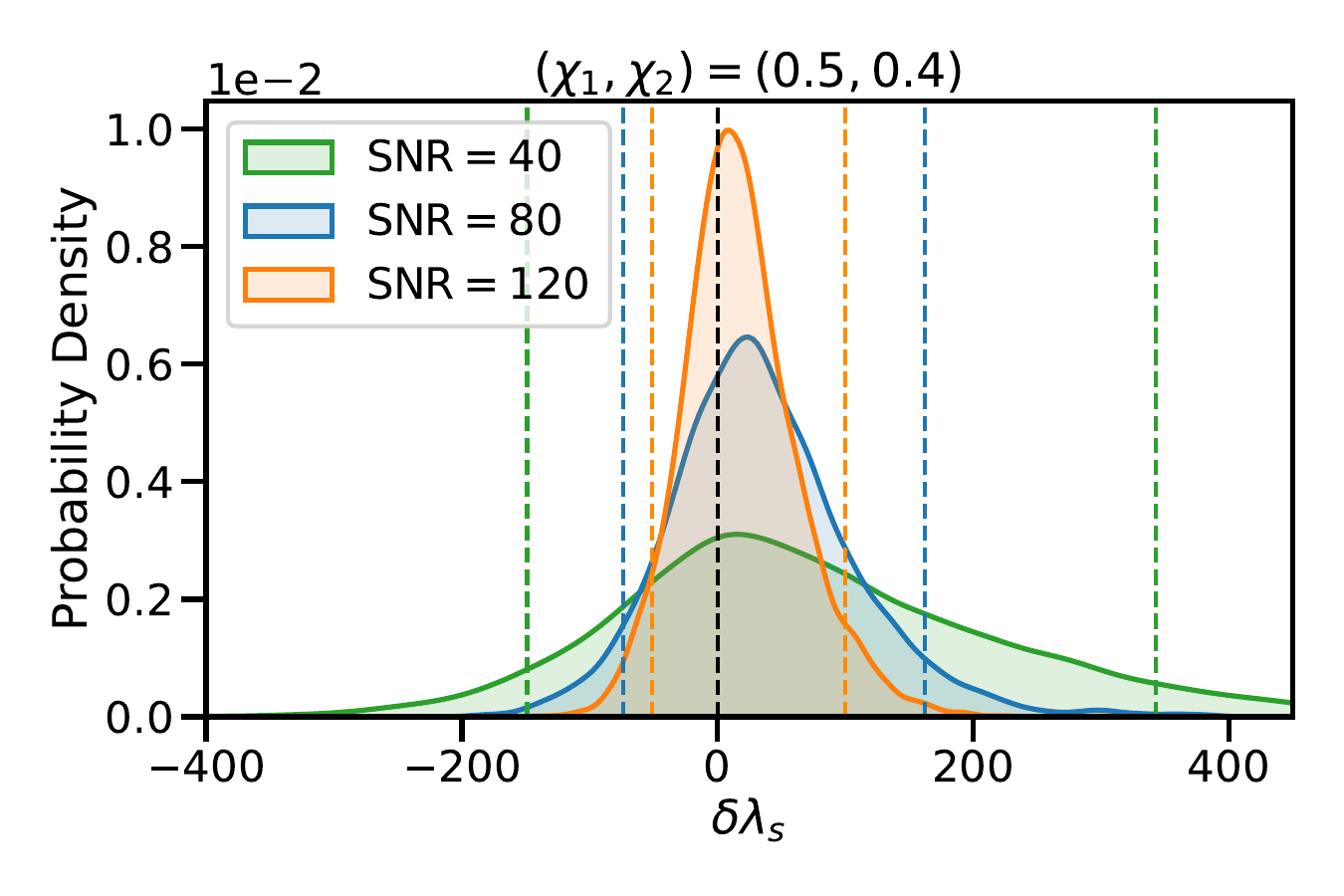}}\vspace{-0.7cm}

\subfigure{\includegraphics[width=0.49\textwidth]{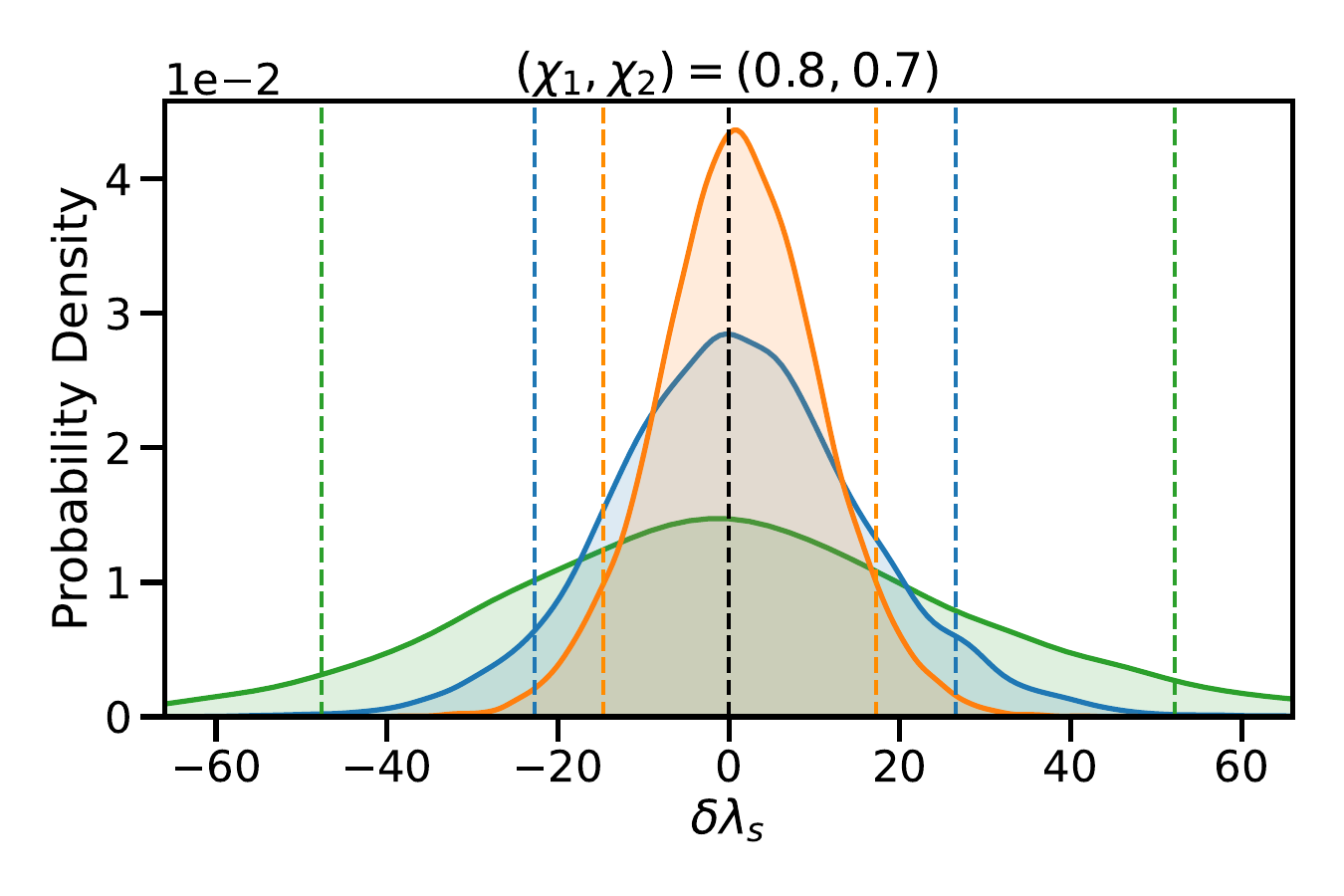}}
\caption{Posterior probability distribution for three SNRs: $40,80,120$. The total mass and mass ratio are fixed to be $40\text{M}_{\odot}$ and $1.2$. Other parameters are same as Fig.~\ref{fig:mass}.}
\label{fig:SNR}
\end{figure}

\begin{figure}[t]
    \centering
    \includegraphics[width=0.49\textwidth]{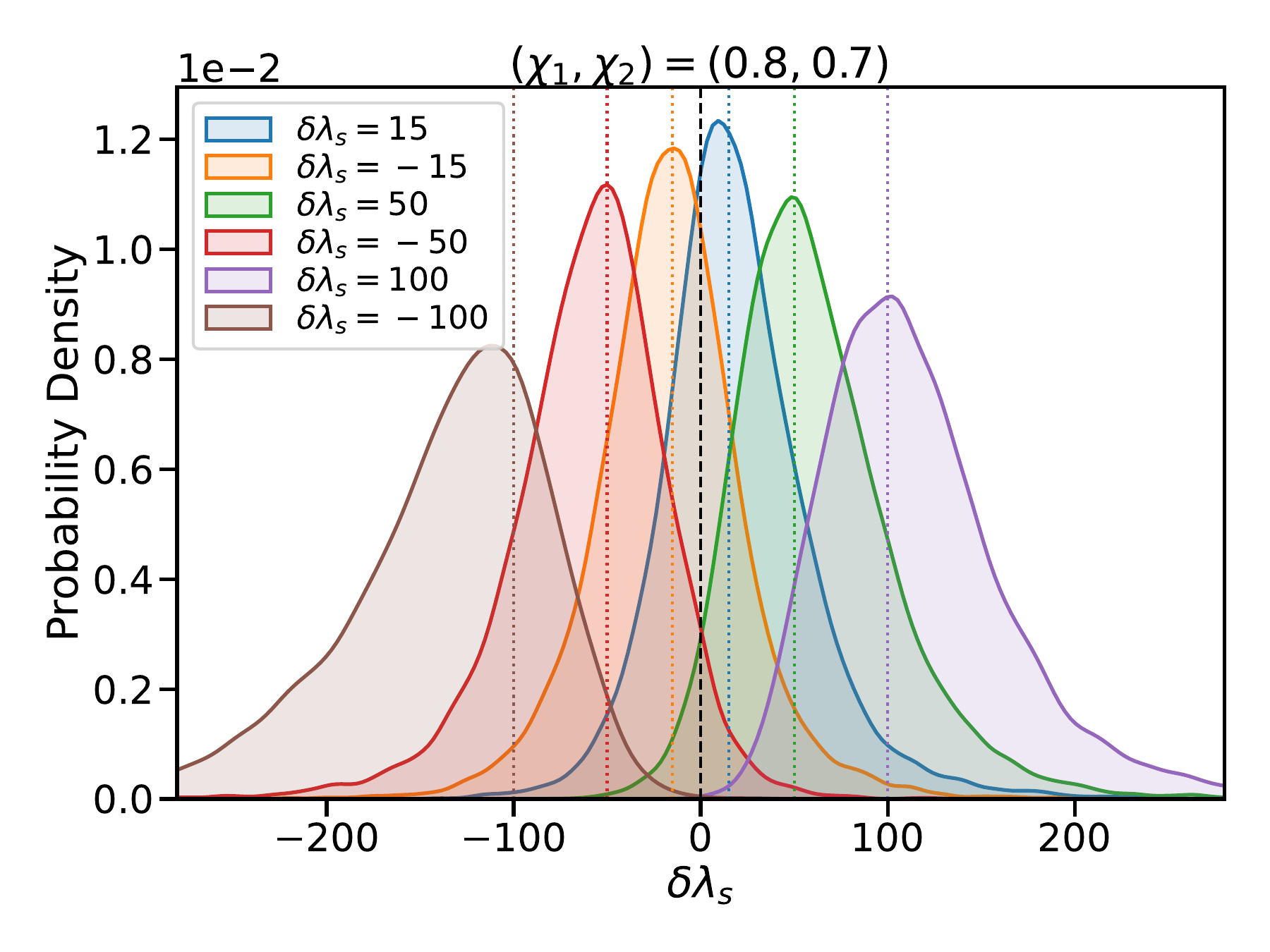}
    \caption{Posterior distributions on $\delta\lambda_s$ for different values of $\delta\lambda_s$. The total mass, mass ratio, and SNR are fixed to be $\text{M} = 40\text{M}_{\odot}$, $\text{q=1.2}$, and $40$, respectively. The dimensionless spins are chosen to be $(\chi_1, \chi_2)=(0.8,0.7)$. Other parameters are the same as Fig.~\ref{fig:mass}.}
    \label{fig:different lambdas}
\end{figure}

\begin{figure}
\centering
   \subfigure{\includegraphics[width=0.45\textwidth]{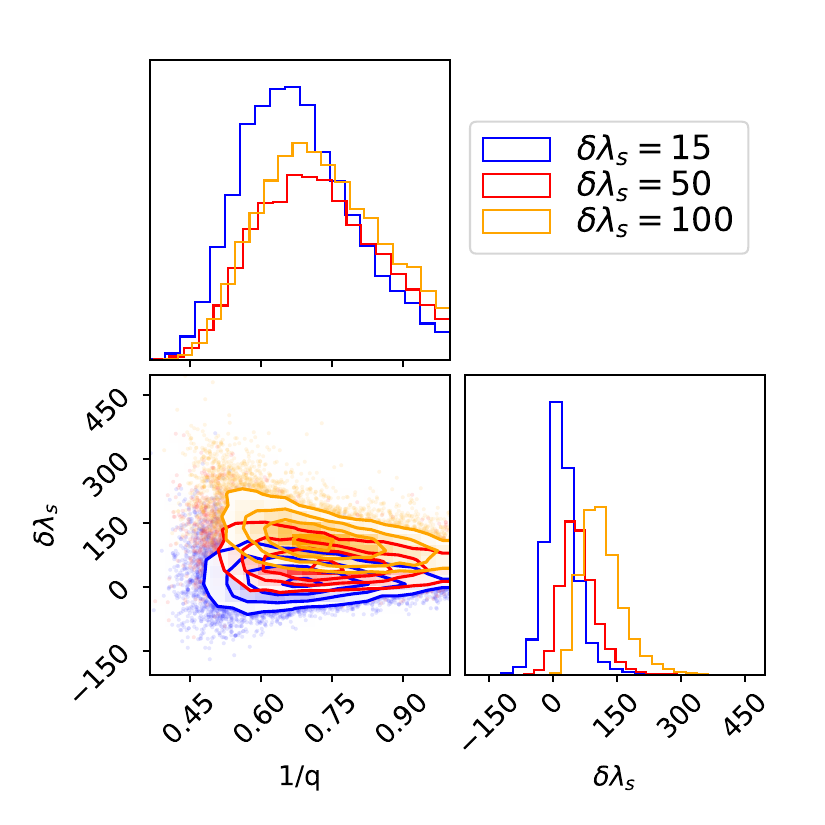}}\vspace{-0.72cm}
   \subfigure{\includegraphics[width=0.45\textwidth]
   {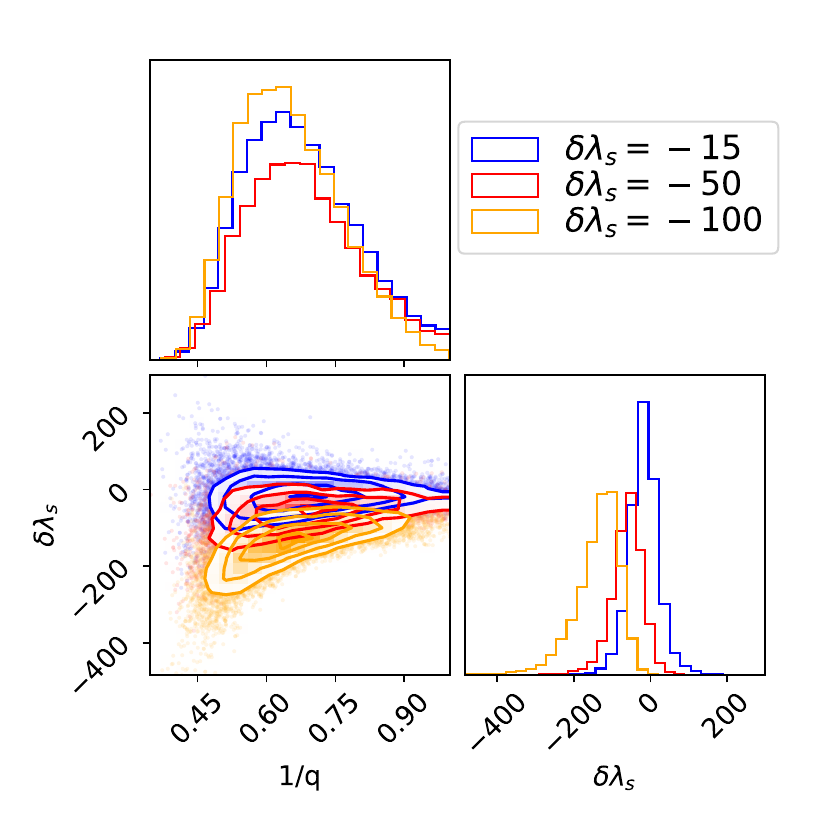}}
\caption{Corner plots for $\delta\lambda_s$ and $\text{q}$ for positive (top panel) and negative (bottom panel) values of $\delta\lambda_s$ injection. The parameters are the same as Fig.~\ref{fig:different lambdas}. As the value of $\delta\lambda_s$ increases, the correlation between $\delta\lambda_s$ and $\text{q}$ increases. The correlations are opposite for positive and negative values of $\delta\lambda_s$ leading to skewness in the $\delta\lambda_s$ posterior.} 
\label{fig:corner}
\end{figure}

\begin{figure*}
\centering
   \begin{subfigure}{\includegraphics[width=0.49\textwidth]{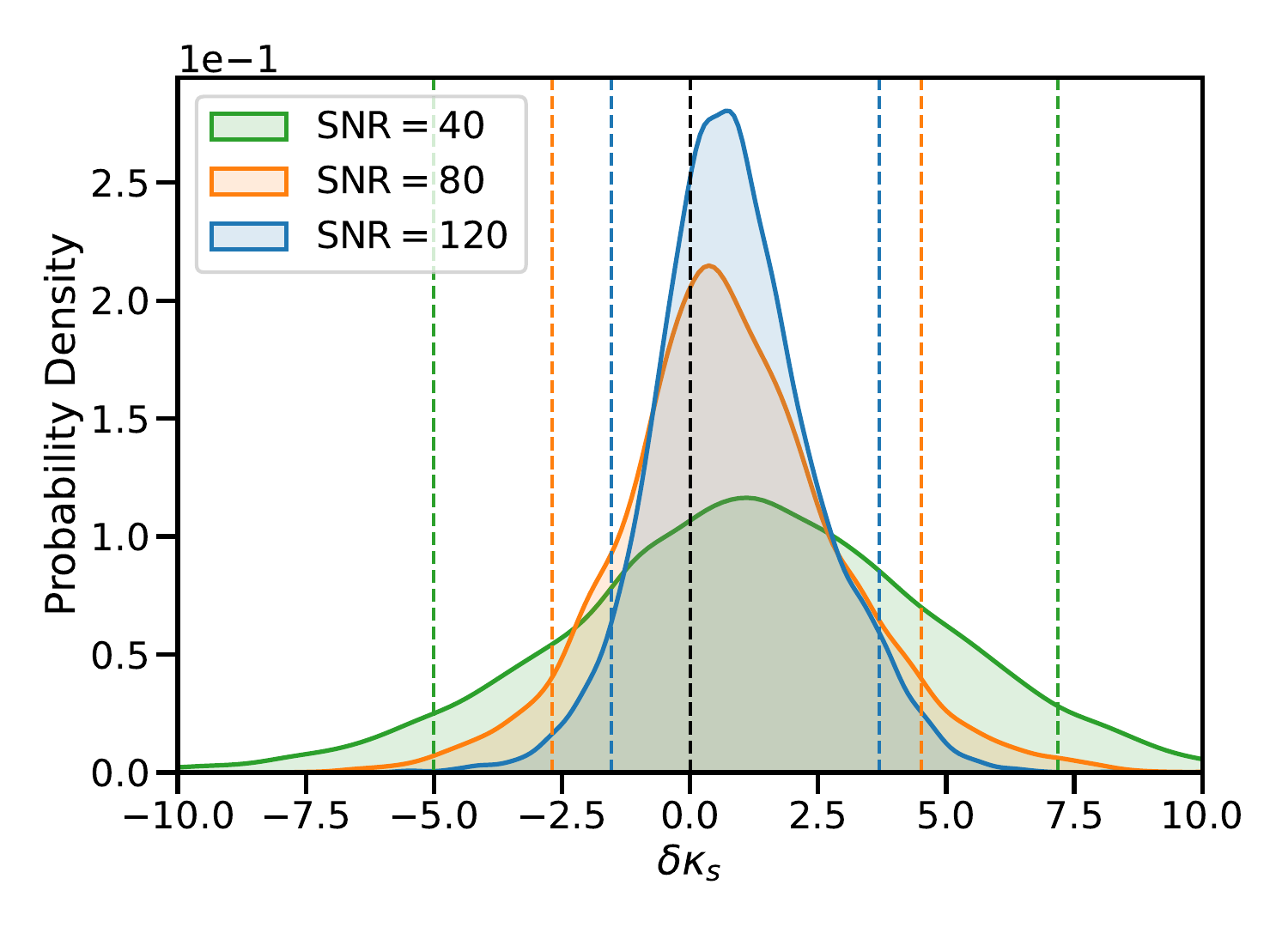}}
   \end{subfigure}
   \begin{subfigure}{\includegraphics[width=0.49\textwidth]{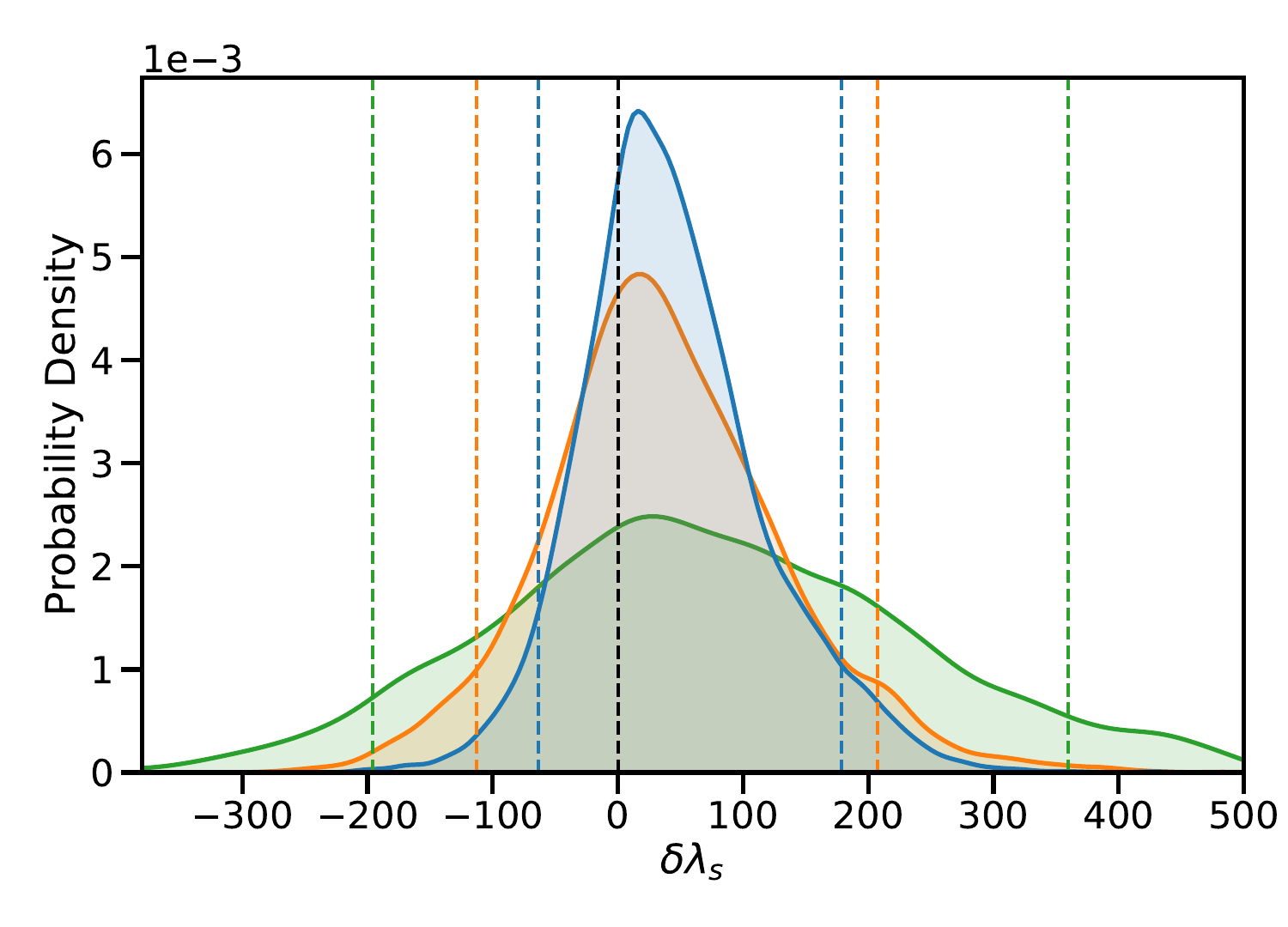}}    
\end{subfigure}
\vspace{-0.4cm}

\caption{Posterior probability distribution on $\delta\lambda_s$ and $\delta\kappa_s$ when measured simultaneously. Total mass is fixed to be $\text{M}=40\text{M}_\odot$ and the mass ratio is $\text{q}=1.2$. Spins are fixed to be ($0.8, 0.7$).}
\label{fig:simultaneous}
\end{figure*}

The main results from the simulated binaries are shown in Fig.~\ref{fig:mass}, Fig.~\ref{fig:massratio}, and Fig.~\ref{fig:SNR}. These figures show the effect of {\it total mass}, {\it mass ratio}, and {\it SNRs} on $\delta\lambda_s$ posterior for three combinations of spin magnitudes. In Fig.~\ref{fig:different lambdas} and Fig.~\ref{fig:corner}, we show the possibility of detecting the non-BBH signature present in the signal for different values of $\delta\lambda_s$. Finally, in Fig.~\ref{fig:simultaneous}, we show the simultaneous bounds on $\delta\kappa_s$ and $\delta\lambda_s$. Now let us examine these figures in detail.

Figure~\ref{fig:mass} shows the posterior probability distribution on $\delta\lambda_s$ for three spin combinations $(\chi_1, \chi_2)$: $(0.3,0.2)$, $(0.5,0.4)$, $(0.8,0.7)$. For each spin combination, different colors show the $\delta\lambda_s$ posterior for three different total masses: $20\text{M}_{\odot}$, $40\text{M}_{\odot}$, and $70\text{M}_{\odot}$.  The mass ratio is fixed to be $1.2$ for all systems. The vertical lines in respective colors represent the $90\%$ credible interval. The black vertical line denotes the injected BBH value $(\delta\lambda_s=0)$. Note that the SNR is fixed at $40$ for these systems. It is evident from the figure that the posteriors become more constraining as the spin magnitudes increase. This is expected because the spin-induced octupole moment is proportional to the cubic in spin terms [see Eq.~\eqref{octupole}], and a larger value of spins increases the contribution of the spin-induced octupole moment in the waveform. Hence, higher spin values lead to better constraints on $\delta\lambda_s$. The posteriors for $\text{M}= 20 \text{M}_{\odot}$ and $\text{M}= 40 \text{M}_{\odot}$ are similar and better constrained compared to $\text{M}= 70 \text{M}_{\odot}$. The better constraints on lower mass systems can be attributed to the more number of GW cycles in the frequency-band of the detector. For $\text{M}= 40 \text{M}_{\odot}$ with $(\chi_1, \chi_2)=(0.8,0.7)$, the $90\%$ credible bounds on $\delta\lambda_s$ are $[-55, 64]$.

In Fig.~\ref{fig:massratio}, we show the $\delta\lambda_s$ posterior for different mass ratios: $1.2$, $3$, and $5$. The total mass and SNR are fixed to be $40 \text{M}_{\odot}$ and $40$, respectively. The posterior on $\delta\lambda_s$ is poorly constrained for the higher mass ratio systems compared to the lower mass ratio systems. Note that the $\delta\lambda_s$ posterior for $\text{q}=1.2$ with $(\chi_1, \chi_2) =(0.5,0.4)$ is better constrained compared to the posterior for $\text{q}=3$ and $\text{q}=5$ with $(\chi_1, \chi_2) =(0.8,0.7)$. The mass ratio trend is consistent with the spin-induced quadrupole moment measurement~\cite{Krishnendu:2017shb} and can be attributed to the correlations among $\delta\lambda_s$, mass ratio, and spins in the waveform. 

Figure~\ref{fig:SNR} shows the effect of SNR on $\delta\lambda_s$ posterior for fixed total mass ($\text{M}=40\text{M}_{\odot}$) and mass ratio ($\text{q}=1.2$). The SNR is varied by varying the luminosity distance to the source. For all spin configurations, the posteriors become narrower as the SNR increases. With an SNR of $120$, the $\delta\lambda_s$ posterior is constrained with good accuracy even for the moderate spins $(\chi_1, \chi_2) = (0.5, 0.4)$. The $\delta\lambda_s$ posterior for $\text{SNR}=120$ with high spins $(\chi_1, \chi_2) = (0.8, 0.7)$ is well constrained within $[-14, 17]$ at $90\%$ credibility. The constraints with high SNR systems give an idea of how the posteriors on $\delta\lambda_s$ would look like for 3G detectors such as CE~\cite{LIGOScientific:2016wof,Reitze:2019iox} and ET~\cite{Sathyaprakash:2011bh} which will observe events with high SNR. Moreover, the better low-frequency sensitivity of CE and ET would further improve the $\delta\lambda_s$ constraints.

\begin{figure}[b]
    \centering
    \includegraphics[width=0.49\textwidth]{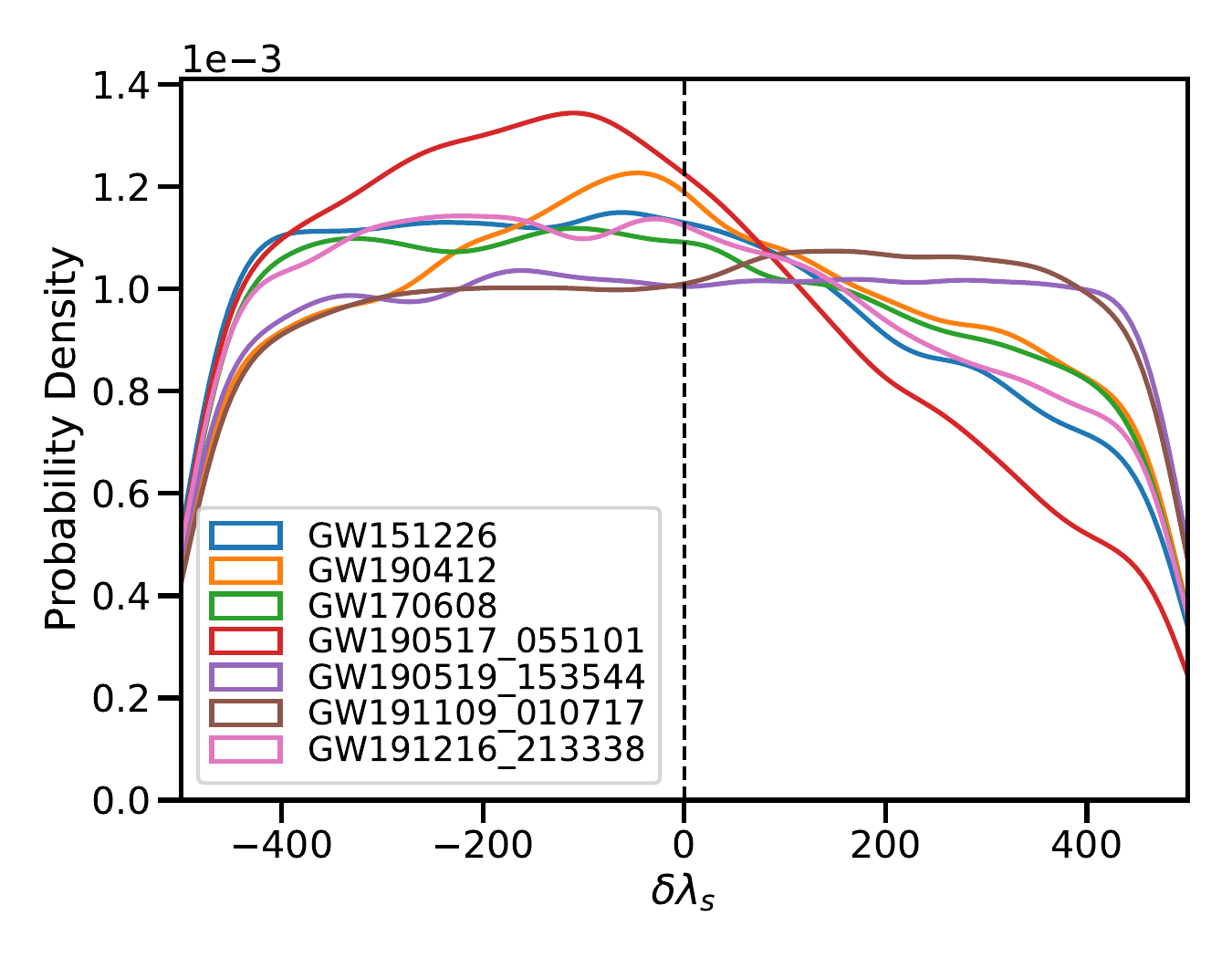}
    \caption{Posterior probability distributions on $\delta\lambda_s$ for selected GWTC-3 events. The posteriors are obtained from the Bayesian analysis of publicly available GWTC-3 data using {\tt IMRPhenomPv2} waveform approximant with $\delta\lambda_s$ corrections in the inspiral part of the waveform.}
    \label{fig:real events}
\end{figure}

\subsection{Detectability of the non-BBH nature from $\delta\lambda_s$ measurements}
In the previous section, we injected a BBH signal ($\delta\lambda_s=0$) and obtained constraints on $\delta\lambda_s$. In this section, we inject a non-BBH signal ($\delta\lambda_s\neq0$) to examine the ability of the HLV network to recover the injected values of $\delta\lambda_s$. In Fig.~\ref{fig:different lambdas}, we show the posteriors corresponding to different injected value of $\delta\lambda_s$ for $\text{M}=40\text{M}_\odot$ system with $\text{q}=1.2$ and $(\chi_1, \chi_2) = (0.8,0.7)$. The $d_L$ is chosen such that network SNR is $40$. The $\delta\lambda_s$ posteriors peak around their injected values. However, in order to rule out zero at $90\%$ credible level and identify the existence of non-BBH characteristics, it requires $|\delta\lambda_s| \gtrsim 50$. We see that smaller values of $\delta\lambda_s$ are better measured compared to the larger values. Also, note that the posteriors for positive (negative) values of $\delta\lambda_s$ show skewness towards the right (left). For larger values of $\delta\lambda_s$, the posteriors show larger skewness.

This trend can be explained by the correlation between $\delta\lambda_s$ and $\text{q}$. Figure~\ref{fig:corner} shows the corner plot for $\delta\lambda_s$ and $\text{q}$. The top and bottom panels are for the positive and negative values of $\delta\lambda_s$, respectively. The correlation between $\delta\lambda_s$ and $\text{q}$ increases as the injected $\delta\lambda_s$  value increases. Moreover, the correlations for positive and negative values of $\delta\lambda_s$ are opposite. The behavior of $\delta\lambda_s$ posteriors in Fig.~\ref{fig:different lambdas} can be associated with the correlations of $\delta\lambda_s$ and $\text{q}$ as seen in Fig.~\ref{fig:corner}.

\subsection{Simultaneous measurement of $\delta\lambda_s$ and $\delta\kappa_s$}
Till now, we have measured $\delta\lambda_s$ by fixing $\delta\kappa_s$ to zero. Ideally, both $\delta\lambda_s$ and $\delta\kappa_s$ should be measured simultaneously to know the true nature of the binary. In this section, we explore the ability of the HLV network to measure both $\delta\kappa_s$ and $\delta\lambda_s$ simultaneously. Figure~\ref{fig:simultaneous} shows the simultaneous measurement of $\delta\kappa_s$ (left panel) and $\delta\lambda_s$ (right panel). The total mass, mass ratio, and spins are fixed to be $\text{M}=40\text{M}_{\odot}$, $\text{q}=1.2$, and $(\chi_1, \chi_2)= (0.8,0.7)$. Three colors refer to three different SNRs: $40$, $80$, and $120$. As expected, when measured simultaneously, the posteriors for both $\delta\kappa_s$ and $\delta\lambda_s$ are wider than the posteriors when both are measured individually. Note that $\delta\kappa_s$ posteriors are $\sim 40 \mbox{--} 60$ times more constrained than $\delta\lambda_s$ posterior. However, as the SNR increases, both $\delta\kappa_s$ and $\delta\lambda_s$ posteriors become more constrained. For $\text{network SNR}=120$, the $90\%$ credible bounds on $\delta\kappa_s$ and $\delta\lambda_s$ are $[-1.5, 3.7]$ and $[-63, 178]$, respectively. HLV network shows promising results for constraining the spin-induced quadrupole and octupole moments together.

\section{Constraints from the third gravitational wave transient catalog}
\label{sec:real event results}
Having studied the behavior of spin-induced octupole moment deformation parameter $\delta\lambda_s$ for various simulated binaries, we report the constraints on $\delta\lambda_s$ from the detected GW events from GWTC-3. We choose events for which the false alarm rates are lower than $10^{-3}$ per year and confidently detected in two or more detectors. Apart from this, we make sure that the inspiral phase $\text{SNR}\geq6$, and there is evidence for non-zero spins. We follow the same event selection criteria as that of Refs. ~\cite{LVK_GWTC3TGR:2021sio, LV_GWTC2TGR:2020tif}. These events also contribute most of their SNR in the inspiral (low-frequency) regime. By adhering to these conditions, we analyze GW151226~\cite{LIGOScientific:2016sjg}, GW190412~\cite{LIGOScientific:2020stg}, GW170608~\cite{LIGOScientific:2017vox}, GW190517\_055101, GW190519\_153544, GW191109\_010717, and GW191216\_213338~\cite{LIGOScientific:2021djp}. We use the publicly available GWTC-3 data for this analysis and employ {\tt IMRPhenomPv2} waveform approximant with $\delta\lambda_s$ corrections in the inspiral part. We obtain the posterior probability distribution on $\delta\lambda_s$ by the Bayesian parameter inference. The prior on $\delta\lambda_s$ is uniform in the range $[-500, 500]$.

Figure~\ref{fig:real events} shows the posterior probability distribution on $\delta\lambda_s$ for selected GWTC-3 events. The posterior distributions for $\delta\lambda_s$ tend to provide limited information. The posteriors for GW190412 and GW190517 show support towards zero. The poor constraints on $\delta\lambda_s$ can be attributed to the low spins of these events~\cite{GWTC1-catalog, GWTC2-catalog, LIGOScientific:2021djp}. From Eq.~\eqref{octupole}, it is clear that the spin-induced octupole moment vanishes for zero spins irrespective of the value of $\lambda$. Moreover, these events have low SNR $\mathcal{O}(10\mbox{--}20)$ with the current sensitivity of GW detectors. However, future 3G detectors, which are expected to have $\mathcal{O}(10)$ fold improved sensitivity and longer inspiral phases due to enhanced low-frequency sensitivity, will enable us to impose stringent constraints on $\delta\lambda_s$. 

\section{Summary and outlook}
\label{sec:conclusions}
In this work, we have developed a Bayesian framework to constrain the nature of compact object binaries based on the measurement of spin-induced octupole moment of binary constituents. We have studied the expected bounds on spin-induced octupole moment deformation parameter $(\delta\lambda_s)$ for various masses, mass ratios, spins, and SNRs. We found that the $\delta\lambda_s$ posteriors are better constrained for binaries with high spins ($\chi_1, \chi_2) = (0.8,0.7$). Moreover, the constraints are generally better for lower mass systems compared to higher mass systems. Furthermore, the lower mass ratio (comparable mass) systems provide better constraints compared to the higher mass ratio systems. For a total mass of $40 \text{M}_{\odot}$ with $\text{q}=1.2$, $(\chi_1, \chi_2) =(0.8,0.7)$, and network SNR of $40$, the $90\%$ credible bounds on $\delta\lambda_s$ are $[-55, 64]$. The constraint on $\delta\lambda_s$ reduces to $[-14, 17]$ for a network SNR of $120$.

We injected different values of $\delta\lambda_s$ to see the detectability of the non-BBH nature present in the GW data. For the highly spinning  $(\chi_1, \chi_2) = (0.8, 0.7)$ system, it requires $|\delta\lambda_s| \gtrsim 50$ for the $\delta\lambda_s$ posterior to exclude zero at $90\%$ credible level and reveal the non-BBH nature of compact object binaries.

Finally, we explored the possibility of the simultaneous measurement of spin-induced quadrupole and octupole moment parameters with the HLV network. Overall, for spins $(\chi_1, \chi_2) = (0.8, 0.7)$ and an SNR of $120$, the HLV network showcases its potential to provide meaningful insights into the nature of compact objects by measuring spin-induced quadrupole and octupole moments together. 

We obtain the first constraints on the $\delta\lambda_s$ parameter from detected binary black hole events~\cite{LIGOScientific:2021djp}. 
The $\delta\lambda_s$ posteriors provide limited information due to low spins of detected GW events and limited sensitivity of current GW detectors. In the future, 3G detectors will be able to put stringent constraints on the nature of compact object binaries by precisely measuring their spin-induced multipole moments. Moreover, combining the bounds from multiple events will improve the combined bound by $\sim 1/\sqrt{N}$, where $N$ is the number of events. Furthermore, implementing the spin-induced multipole moment corrections to a more sophisticated waveform with higher modes and precession may further improve the bounds on spin-induced multipole moments. These possibilities can be explored in a future study.

\section*{Acknowledgments}  
We thank K.~G.~Arun for useful discussions and comments on the draft. It is a pleasure to thank Anuradha Gupta for the careful reading of the manuscript. Authors also thank B.~S.~Sathyaprakash, Nathan Johnson-McDaniel, Parthapratim Mahapatra, Sajad A. Bhat, Sayantani Datta, and Poulami Dutta Roy for valuable discussions and/or comments on the draft. P.S. acknowledges partial support from the Infosys Foundation. N.V.K. acknowledges the support from Science and Engineering Research Board (SERB) for the National postdoctoral fellowship (Reg. No. PDF/2022/000379). Computations were performed on the CIT cluster provided by the LIGO Laboratory. We acknowledge National Science Foundation Grants No. PHY-0757058 and No. PHY-0823459. This material is based upon work supported by NSF's LIGO Laboratory which is a major facility fully funded by the National Science Foundation. This document has LIGO preprint number {\tt LIGO-P2300231}. We used the following software packages:
{\tt LALSuite}~\cite{lalsuite}, {\tt bilby}~\cite{Ashton:2018jfp}, {\tt bilby\textunderscore pipe}~\cite{Romero-Shaw:2020owr}, {\tt NumPy}~\cite{2020Natur.585..357H}, {\tt PESummary}~\cite{Hoy:2020vys}, {\tt Matplotlib}~\cite{2007CSE.....9...90H}, {\tt Seaborn}~\cite{Waskom2021}, {\tt jupyter}~\cite{soton403913}, {\tt dynesty}~\cite{speagle2020dynesty}, {\tt corner}~\cite{corner}. 

\bibliographystyle{apsrev}
\bibliography{reference}
\end{document}